\documentclass[12pt]{spieman}  
\usepackage{amsmath,amsfonts,amssymb}
\usepackage{graphicx}
\usepackage{setspace}
\usepackage{tocloft}

\usepackage[mathlines]{lineno}
\usepackage{graphicx}
\usepackage{color}

\usepackage{amssymb} 
\usepackage{amsmath} 
\usepackage{bm} 
\usepackage{upgreek}
\usepackage{longtable}

\newcommand{\ve}[1]{\mathbf{#1}} 
\newcommand{\ma}[1]{\mathsf{#1}} 
\newcommand{\imu}{i} 
\newcommand{\Rtwo}{{\mathbb{R}^2}}

\title{Detective quantum efficiency of photon-counting CdTe and Si detectors for computed tomography: a simulation study}

\author[a,b,*]{Mats Persson}
\author[b]{Adam Wang}
\author[a,b,c]{Norbert J. Pelc}
\affil[a]{Department of Bioengineering, Stanford University, Stanford, CA 94305, USA}
\affil[b]{Department of Radiology, Stanford University, Stanford, CA 94305, USA}
\affil[c]{Department of Electrical Engineering, Stanford University, Stanford, CA 94305, USA}

\cftpagenumbersoff{figure}
\cftpagenumbersoff{table} 
\begin{document} 
\maketitle

\begin{abstract}

\textbf{Purpose:} Developing photon-counting CT detectors requires understanding the impact of parameters such as converter material, absorption length and pixel size. We apply a novel linear-systems framework, incorporating spatial and energy resolution, to study realistic silicon (Si) and cadmium telluride (CdTe) detectors at low count rate. 

\textbf{Approach:} We compared CdTe detector designs with $0.5\times0.5\; \mathrm{mm}^2$ and $0.225\times0.225\; \mathrm{mm}^2$ pixels and Si detector designs with $0.5\times0.5\; \mathrm{mm}^2$ pixels of 30 and 60 mm active absorption length, with and without tungsten scatter blockers. Monte-Carlo simulations of photon transport were used together with Gaussian charge sharing models fitted to published data. 

\textbf{Results:} For detection in a 300 mm thick object at 120 kVp, the 0.5 mm and 0.225 mm pixel CdTe systems have 28-41 $\%$ and 5-29 $\%$ higher DQE, respectively, than the 60 mm Si system with tungsten, whereas the corresponding numbers for two-material decomposition are 2 $\%$ lower to 11 $\%$ higher DQE and 31-54 $\%$ lower DQE compared to Si. We also show that combining these detectors with dual-spectrum acquisition is beneficial. 

\textbf{Conclusions:} In the low-count-rate regime, CdTe detector systems outperform the Si systems for detection tasks, while silicon outperforms one or both of the CdTe systems for material decomposition. \\

\end{abstract}

\keywords{x-ray computed tomography, photon-counting, silicon detector, cadmium telluride detector, detective quantum efficiency, performance comparison}

{\noindent \footnotesize\textbf{*}Mats Persson, present affiliation: KTH Royal Institute of Technology, Stockholm, Sweden. \linkable{mats.persson@mi.physics.kth.se} }

\begin{spacing}{2}   


\section{Introduction}
\label{sec:introduction}
Photon-counting detectors are expected to become the next major advance in x-ray computed tomography (CT).\cite{taguchi_vision_2020,giersch_energy_weighting,shikhaliev_first_results} Whereas the energy-integrating detectors in use today measure the total incident energy in the x-ray beam during each measurement time interval, photon-counting detectors are able to count the individual photons and measure their energy. This enables improved signal-to-noise ratio. Moreover, the energy-resolving capabilities can allow spectral imaging with better energy separation than what is practically achievable with existing dual-energy CT technologies such as dual source\cite{johnson_dual_source_dual_energy}, kVp-switching\cite{li_quantification_of_ctdivol}, split filtration\cite{euler_split_fllter} and dual layer\cite{sauter_dual_layer} detectors. Photon-counting detectors can also provide higher spatial resolution.

The most frequently considered detector materials for photon-counting CT are cadmium telluride (CdTe), cadmium zinc telluride (CZT) and silicon (Si). A large amount of research focused on CdTe/CZT\cite{schlomka_experimental_feasibility,yu_mayo_research_prototype_evaluation,ronaldson_toward_quantifying} since these materials have high attenuation in the diagnostic x-ray energy range (30-150 keV) and a high fraction of photoelectric absorption. However, a drawback of these materials is the high probability of K-fluorescence, which can cause part of the energy deposited in an interaction to be emitted as a fluorescence photon and reabsorbed in another detector pixel or to escape. Silicon,\cite{bornefalk_silicon_strip_pmb, persson_energy_resolved_ct_imaging} on the other hand, has low attenuation meaning that a large absorption length (on the order of 30-60 mm) is needed to get a high detection efficiency. Another drawback of silicon is the high fraction of Compton scatter, a process in which an incident photon deposits a fraction of its energy at the original site of incidence and subsequently either leaves the sensor or causes one or more additional interactions in other parts of the sensor material. However, compared to CdTe, silicon suffers less from K-fluorescence and has shorter charge collection time and therefore better capability to measure high x-ray fluence rates. Silicon detectors can also easily be made with several depth segments, further improving their count rate capability. Additionally, both silicon and CdTe and CZT suffer from charge sharing, which occurs when a photon interacts close to a pixel border and the deposited energy is divided between two or more detector pixels.\cite{xu_charge_sharing,taguchi_spatioenergetic} K-fluorescence, Compton scatter and charge sharing all lead to spatial blurring, lower signal-to-noise ratio and degraded energy resolution.

The effort to develop improved photon-counting CT detectors raises an important question: how should a photon-counting detector be designed in order to maximize its performance? X-ray detector performance is frequently measured by the detective quantum efficiency (DQE).\cite{icru_54,cunningham_linear_systems_chapter} This is a number between zero and one, usually reported as a function of spatial frequency, which measures the dose efficiency of the studied detector relative to an ideal detector with perfect absorption efficiency and spatial resolution.

Previous authors have used the frequency-dependent DQE to study non-energy-resolving photon-counting detectors\cite{acciavatti_maidment_phc_ei,tanguay_dqe_2013,xu_cascaded_sys_phc, stierstorfer_dqe}, and investigated the zero-frequency (large-area) spectral imaging performance \cite{chen_optimization_of_beam_q, cammin_evaluation_spectral_dist, taguchi_energy_response_pileup, tanguay_dqe_2015, faby_statistical_correlations, taguchi_spatioenergetic, rajbhandary_spatioenergy_spie,taguchi_pctk}. For a more complete characterization of spectral x-ray detectors, however, the spatial-frequency dependence and the impact of energy resolution must be studied together. Steps in this direction have been taken for dual-energy x-ray systems\cite{richard_noise_reduction_dual_energy} and for photon-counting detectors \cite{fredenberg_contrast_enhanced_spectral,yveborg_theoretical_comp,chen_size_dependent_parameters}.

The conventional DQE measure only applies to non-energy-resolving detectors and thus does not take into account that improved performance can be attained through optimal weighting of different x-ray energies.\cite{cahn_dqe, TGS_image_based} In a previous publication, we described a theoretical framework for addressing such questions, based on linear-systems theory, and showed how a generalized DQE measure can be defined for characterizing performance for both feature detection and material quantification tasks.\cite{persson_framework_performance_characterization} In a feature detection task, the objective is to determine whether a feature of specified frequency content and material composition is present or absent in an image. A quantification task, on the other hand, is based on the observation that the x-ray attenuation of any substance in the human body can be well approximated by a linear combination of two basis materials, or three if a K-edge contrast agent such as iodine is present.\cite{alvarez_macovski,roessl_proksa_k_edge} The objective for a quantification task is then to measure the amount of a given basis material located between source and detector. This measure can be viewed as a spatial-frequency-dependent version of the Cram\'{e}r-Rao lower bound that is commonly used to assess material decomposition performance\cite{roessl_hermann_crlb}. An equivalent description of quantification performance, but in the spatial domain rather than the frequency domain, has been published in Ref. \citenum{rajbhandary_frequency_dependent_dqe}.


The purpose of the present investigation is to use this novel framework to compare typical proposed designs for Si and CdTe photon-counting detectors in a simulation study. To this end, we simulate photon transport with Monte-Carlo methods and charge sharing with a charge cloud model fitted to published data. From these simulations, we calculate the task-specific detective quantum efficiencies (DQEs) for several detection and quantification tasks. Preliminary versions of some of our results were presented in Ref. \citenum{persson_spie_detector_comparison}.
\section{Materials and methods}
\label{sec:methods}
\subsection{Detector geometries}

\begin{figure}
  \centering
  \includegraphics[scale=0.8]{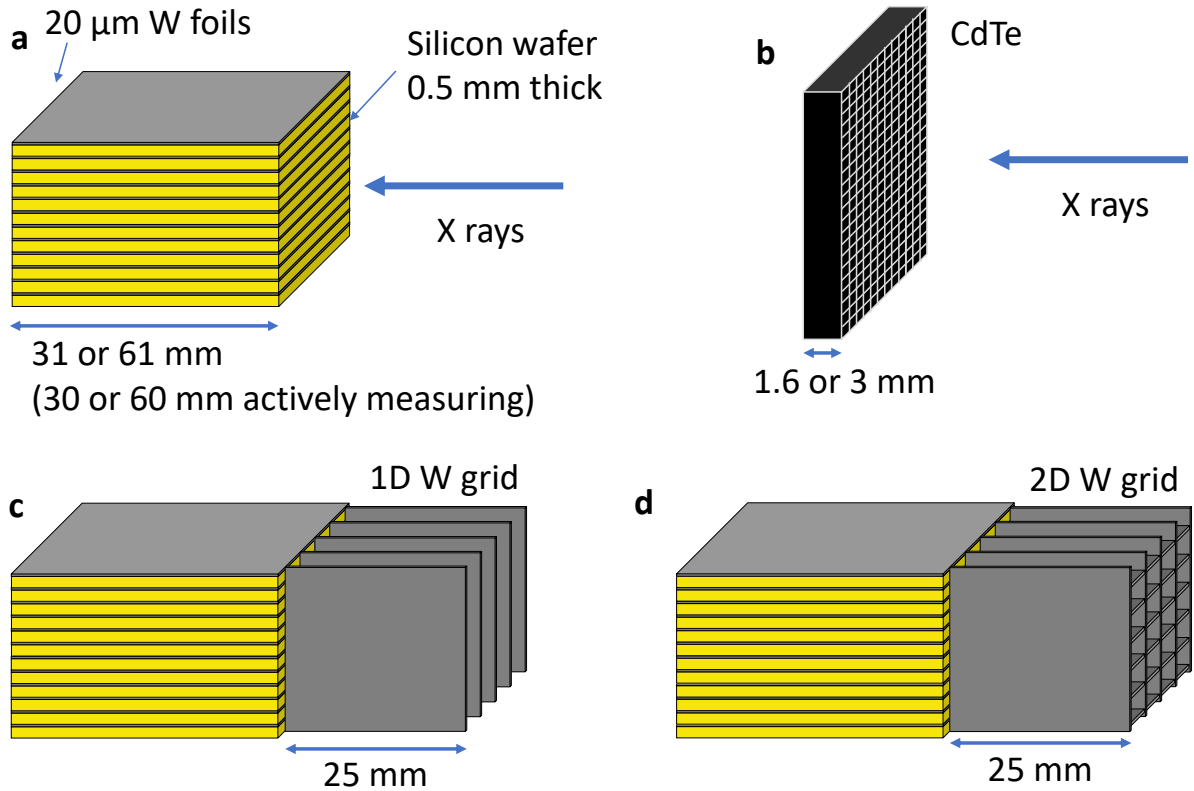}
  \caption[Detector geometries.]{Detector simulation geometries. (a) Silicon detector with tungsten foils. (b) CdTe detector. (c) Silicon detector with 1D anti-scatter grid. (d) Silicon detector with 2D anti-scatter grid. Similar anti-scatter grids are used with the CdTe detector.}
  \label{fig:detector_drawing}
\end{figure}

Our model of a silicon strip detector mimics the detector described in Refs. \citenum{bornefalk_silicon_strip_pmb} and \citenum{xu_silicon_strip_detector}. This detector consists of wafers (0.5 $\times$ 31  $\times$ 100 $\mathrm{mm}^3$), divided into $\mathrm{0.5\times0.5\; \mathrm{mm}^2}$ detector pixels oriented with their edge facing the x-ray source (edge-on), and stacked together to form a volumetric detector with 100\% area coverage (Fig. \ref{fig:detector_drawing}(a)). To explore the effect of absorption length, a design with 61 mm Si length was also simulated. Dead silicon volume, 0.5 mm in the front and 0.5 mm in the back, that does not count photons modeled the guard ring used to prevent current leakage in the silicon wafer. This gives active counting absorption lengths of 30 and 60 mm, respectively. In contrast to the referenced real-world detector design, the simulated detector is not segmented along the depth direction, i.e. along the direction of the x-ray beam (although depth segmentation was implemented for the simulations used to fit the charge cloud model, see Sec. \ref{sec:charge_sharing_model_fitting}). For each of the two designs (30 and 60 mm), we studied two variants: one with and one without 20 ${\upmu}$m tungsten foils interspersed between the silicon wafers, oriented orthogonally to the y axis in our coordinate system. These foils are intended to stop scattered x-ray photons, to decrease the probability that an incident photon is counted more than once\cite{bornefalk_silicon_strip_pmb}. When tungsten foils are introduced, the pixel size decreases from $\mathrm{0.5\times0.5\; \mathrm{mm}^2}$ to $\mathrm{0.5\times0.48\; \mathrm{mm}^2}$ to leave space for the tungsten foils, and the active detector area fraction is decreased from 100\% to 96\%.

Two different CdTe detector designs were studied (Fig. \ref{fig:detector_drawing}(b)). The first has a converter thickness of 3 mm and a pixel size of $0.5\times0.5 \; \mathrm{mm}^2$. The second design has a converter thickness of 1.6 mm and a pixel size of $0.225\times0.225 \; \mathrm{mm}^2$, similar to the detector described in Ref. \citenum{yu_mayo_research_prototype_evaluation}. Designs with even smaller pixels have also been developed\cite{ronaldson_toward_quantifying} but these require charge summing logic which is challenging at the high count rates in clinical CT, and were therefore not included in the present study. 

\subsection{Photon transport simulation}
\label{sec:photon_transport_simulation}
We used Geant4-based GATE\cite{jan_gate} to simulate photon transport. For each energy (20 to 150 keV in steps of 1 keV), a monoenergetic beam of $4 \cdot 10^4$ photons impinging on the detector was simulated. To ensure that no boundary effects affected the simulation, the simulated detector area was $100\times 100 \; \mathrm{mm}^2$ for CdTe and Si, i.e. the Si detector consisted of 201 stacked Si wafers, 100 mm long, with interspersed tungsten foils. Two types of simulations were made, one with the full pixel unit cell illuminated for calculating autocovariance and one with only a sub-rectangle of the pixel illuminated (see Sec. \ref{sec:method_psf}-\ref{sec:method_acf}). The GATE simulation was run with the Penelope physics list, and the minimum range for photons and electrons, determining the low-energy cutoff, was set to $10 \; \upmu \mathrm {m}$ in Si and CdTe and $1 \; \upmu \mathrm {m}$ in the tungsten foils.
\subsection{Charge sharing simulation}
The list of interactions from the Monte-Carlo simulation was used as input to a charge sharing simulation implemented in MATLAB (The MathWorks, Inc. Natick, MA, USA). Instead of simulating the physics of charge transport in the material, which would be computationally demanding, we used an approximative simulation model where each interaction gives rise to a spherically symmetric charge cloud. The registered pulse height in each detector pixel is then assumed to be proportional to the amount of charge in the projection of this charge cloud on the surface of the semiconductor volume belonging to one pixel. The projection is in the direction of the electric field, i.e. parallel to the beam for CdTe and parallel to the wafer normal for Si. The pulse height was also assumed to be independent of the distance between the point of interaction in the direction orthogonal to the pixel electrode. This is a simplified model, but it is computationally inexpensive and can be fitted to measured or simulated data by adapting the charge cloud as a function of the interaction energy.

In a real detector, the spectrum that is actually registered may be different than the spectrum of pulse heights corresponding to the collected charge due to pulse pileup. In this work, however, we make the simplifying assumption that the photon fluence rate is low enough that pileup does not occur.
\subsubsection{Charge sharing model fitting}
\label{sec:charge_sharing_model_fitting}
The model described above was used to perform charge sharing simulations for monochromatic incident beams with different choices of charge cloud sizes (which were taken to be independent of energy in this simulation) and for both Gaussian (standard deviation $\sigma=0-40 \;\upmu \mathrm{m}$ for CdTe and $\sigma=0-35 \;\upmu \mathrm{m}$ for Si, in steps of $1 \; \upmu \mathrm {m}$) and uniform spherical charge clouds ($0-80 \;\upmu \mathrm{m}$ radius for CdTe and $0-60 \;\upmu \mathrm{m}$ for Si, in steps of $1 \; \upmu  \mathrm {m}$). Spectra of registered energies were recorded in steps of 1 keV.

The simulations were fitted to previously published spectral data. For silicon, we used reference spectra from Ref. \citenum{liu_energy_resolution} where monochromatic synchrotron measurements at 40, 60, 80, 100 and 120 keV are detailed (cumulative spectra that were made available to us by the authors of Ref. \citenum{liu_energy_resolution}). To imitate the published measurement, we simulated a rectangular beam covering five pixels in a simulation of a single Si wafer with $0.5\times0.4 \; \mathrm{mm^2}$ pixels in a linear array and 29.7 mm active absorption length, divided into 16 depth segments. Note that this geometry is slightly different from the configurations whose performance we assess below, and was used only to estimate the parameters in the charge sharing model. For each energy, the measurements for three illuminated strips were summed and the resulting cumulative spectrum of registered energies was fitted, with electronic noise as a free parameter, to the experimentally measured data processed in the same way. Measurements with low photon statistics were excluded from the fitting, so that the total number of used depth segments varied between 2 and 16.

For CdTe, we fitted our model to the simulated spectra published in Ref. \citenum{xu_charge_sharing}, where a more detailed simulation including charge transport in the semiconductor material is described. The cited publication includes spectra for three pixel sizes: $0.3\times0.2$, $0.5\times0.4$ and $1\times1\; \mathrm{mm^2}$, for monoenergetic 60 and 100 keV x-ray beams. These were mimicked by simulating a rectangular beam covering one pixel of a two-dimensional 3 mm thick CdTe detector and registering events taking place in any pixel. The resulting cumulative spectrum of registered energies was fitted to the cumulative spectrum obtained from the published data. Since the fitting was made to a noiseless simulated spectrum, electronic noise was not modeled in the fit for CdTe.

After fitting a charge cloud size to each energy data point (and for CdTe, each pixel size), we fit all these charge cloud sizes to a functional relationship between charge cloud size and photon energy. This was done for both the uniform and the Gaussian models, but since previous studies demonstrated that Gaussian models generate realistic spectra\cite{taguchi_pctk}, only the Gaussian charge cloud model was used in subsequent simulations.

Since we did not have access to spectra for a 1.6 mm thick CdTe sensor to which a charge sharing model could be fitted, we simulated the 1.6 mm thick CdTe detector using two different charge cloud models: the one for the 3 mm thick CdTe, and one where the charge cloud was rescaled linearly with thickness, by a factor of 1.6/3. The true charge cloud size for this detector design can be expected to lie somewhere between these two.

\subsubsection{Point-spread function simulation}
\label{sec:method_psf}

Using the list of interactions from Sec. \ref{sec:photon_transport_simulation} and the charge sharing from Sec. \ref{sec:charge_sharing_model_fitting}, a spectrum of detected energies was simulated for each pixel and for each incident energy, in steps of 1 keV. This simulation was performed on a $41\times41$ pixel grid ($20.5\times20.5 \; \mathrm{mm}^2$) for Si in order to capture Compton scattered photons, and a $5\times5$ ($2.5\times2.5 \; \mathrm{mm}^2$) or $11\times11$ pixel grid ($2.48\times2.48 \; \mathrm{mm}^2$) pixel grids for the CdTe detector with 0.5 or 0.225 pixels, respectively, to obtain the point-spread function $h(\ve{r},E)$ giving the expected number of counts registered in each pixel as a function of position $\ve{r}=(x,y)$ and incident x-ray energy $E$. In order to study the point-spread function with higher resolution than that given by the pixel size, nine sub-beams were simulated, each with an area of $\frac{1}{3}\times\frac{1}{3}$ pixels, which together cover the entire pixel. This gives the point-spread function of the system convolved with a two-dimensional rect function with a side of $1/3$ pixel; and this was corrected through deconvolution (see Sec. \ref{sec:dqe_calculation}). The benefit of using a nonzero beam width is that the position dependence of the detector response within the pixel can be taken into account fully without having to use a large number of beams. For the CdTe detector, which is translation invariant in two dimensions, all sub-beams were obtained by translating one single sub-beam photon transport simulation relative to the pixel grid and carrying out the charge-sharing simulation for each translated position. For the Si detector, whose layered structure makes it translation invariant in only one dimension, two different photon transport simulations were made: one in the pixel center and one at the wafer edge. The other sub-beam locations were then obtained by translating these two, before the charge sharing simulation. Finally, the point-spread function was forced to be symmetric by averaging it with its mirror image in each of the $x$ and $y$ dimensions.
\subsubsection{Autocovariance simulation}
\label{sec:method_acf}

The autocovariance of the noise was simulated through a similar procedure as in Sec. \ref{sec:method_psf} but with a square beam covering the entire pixel unit cell. For each \textit{pair} of pixels and deposited energies (discretized in 1 keV steps), the number of photons $N_{\ve{n},m,\ve{n'},m'}$ registered in both were counted. Here, $\ve{n}=(n_x,n_y)$ is a 2D pixel index and $m$ and $m'$ are deposited energy indices. By summing over pairs of pixels with the same relative offset to each other, we obtained the autocovariance $K^{s,\mathrm{ub}}_{\ve{\Delta n},m,m'}=\sum_\ve{n}N_{\ve{n},m,\ve{n}+\ve{\Delta n},m'}$ of the sampled detector signal for a detector under homogeneous full-field illumination. Here, $\mathrm{ub}$ denotes unbinned, i.e. this is the autocovariance before the counts are put into energy bins. This formula is based on the observation that the covariance between two Poisson distributed photon count values is equal to the expected value of the number of counts registered in both. If only pixel $\ve{n}$ had been illuminated, the covariance would therefore have been given by $N_{\ve{n},m,\ve{n}+\ve{\Delta n},m'}$, and the covariance for full-field illumination is obtained by summing this single-pixel illumination formula over different beam locations. As with the point-spread function, the autocovariance was forced to be symmetric in each of the $x$ and $y$ dimensions.
\subsection{DQE calculation}
\label{sec:dqe_calculation}
Our method of calculating detector performance metrics builds on the theoretical framework presented in Ref. \citenum{persson_framework_performance_characterization}, which derives formulas presented here without proof. The simulated point-spread function $h(\ve{r},E)$, was used to calculate the unbinned transfer function that relates $A_l$, the line integral of basis coefficent $l$, to the number of counts at energy discretization point $m$:
\begin{equation}
\label{transfer_matrix}
H_{m,l}^{\mathcal{B},\mathrm{ub}}(\ve{u}) = \int_{E=0}^{E_{\mathrm{max}}} \overline{q}^\mathrm{tot}\left(\frac{\partial \overline{q}(E)}{\partial A_l}/\frac{\partial \overline{q}_{\mathrm{tot}}}{\partial A_l} \right)H_m(\ve{u},E) \mathrm{d}E
\end{equation}

where $\ve{u}$ is spatial frequency, $H_m(\ve{u},E)=\int_{\Rtwo} h_m(\ve{r},E)\mathrm{e}^{-2\pi \imu\ve{u}\cdot\ve{r}} \mathrm{d}\ve{r}$ is the Fourier transform of the point-spread function at discretized deposited energy $m$, $\overline{q}(E)$ is the transmitted x-ray spectrum (photons per area and energy), and $\overline{q}^\mathrm{tot}=\int_0^{E_{\mathrm{max}}}q(E)\mathrm{d}E$ is the total number of photons per area impinging on the detector.

For the set $\mathcal{B}$ of basis materials, we used subsets of $\left\{\right.$water, cortical bone, iodine (10 mg/ml) and gadolinium (1 mg/ml)$\left.\right\}$\cite{nist_xcom} but never all four at the same time. $\overline{q}(E)$ were tungsten anode x-ray spectra at 80, 120 and 140 kVp with $12^\circ$ anode angle (calculated with the TASMICS \cite{hernandez_boone_tasmics} model  implemented using Spektr 3.0) filtered with 3 mm Al and 0.9 mm Ti (giving a half-value layer of 8.26 mm Al at 120 keV) plus varying thicknesses of water representing the patient. The 140 kVp spectrum was additionally filtered with 0.4 mm Sn to simulate the Sn140 kVp beam quality commonly used in DECT\cite{primak_tin_filtration}. 

The number of energy bins was 8 for the Si detector\cite{bornefalk_silicon_strip_pmb,persson_energy_resolved_ct_imaging}, 5 for the CdTe detector with 0.5 mm pixels and 2 for the CdTe detector with 0.225 mm pixels\cite{yu_mayo_research_prototype_evaluation}. 

The transfer function from basis materials to counts in energy bins was calculated as a weighted sum of the original transfer function over energies: $H_{k,l}^\mathcal{B}(\ve{u})= \sum_m  B_{k,m} H_{m,l}^{\mathcal{B},\mathrm{ub}}(\ve{u})$. The bin response functions $B_{m,k}$ were rectangular functions equal to one for energy points $m$ located with energy bin $k$ and 0 otherwise, smoothed by convolution with a Gaussian kernel to model electronic readout noise. The kernel RMS was $1.6 \; \mathrm{keV}$\cite{liu_energy_resolution} ($3.8 \; \mathrm{keV}$ FWHM) for Si and 3.0 keV ($7.0 \; \mathrm{keV}$ FWHM) for CdTe\cite{mccollough_cern_2017,barber_optimizing_cdte}.

The unbinned autocovariance $K^{s,\mathrm{ub}}_{\ve{\Delta n},m,m'}$ can similarly be summed over energy bins to obtain the autocovariance for energy bin counts: 
\begin{equation}
  K^{s}_{\ve{\Delta n},k,k'}=\sum_m \sum_{m'} B_{k,m} B_{k',m'}  K^{s,\mathrm{ub}}_{\ve{\Delta n},m,m'}, \;\;\;\;\; \ve{\Delta n} \ne 0.
\end{equation}
Since each interaction is counted only once, electronic noise can never lead to correlation between different energy bins in the same detector channel, i.e. its only effect on the autocovariance matrix for $\ve{\Delta n}=0$ is to blur the energy spectrum on its diagonal. Therefore, the $\ve{\Delta n}=0$ entries must be treated in a different way from the other entries: 

\begin{equation}
  K^s_{\ve{0},k,k'}=\left\{
  \begin{matrix}
    \sum_m B_{k,m} K^{s,\mathrm{ub}}_{\ve{0},m,m}, &k = k' \\
    0 & k  \ne k'
  \end{matrix}
  \right.
\end{equation}
where we have made the assumption that $K^{s,\mathrm{ub}}_{\ve{0},m,m}$ is diagonal, i.e. the detector never counts the same photon twice in the same readout channel.

The cross-spectral density can then be obtained from the autocovariance as

\begin{equation}
  \label{eq:csd}
  \left[W_{d^+}(\ve{u})\right]_{k,k'}=\frac{1}{\Delta_x \Delta_y}\sum_{\Delta \ve{n}=-\infty}^{\infty}K^s_{\ve{\Delta n},k,k'}\mathrm{e}^{-2\pi \imu(\ve{u}\cdot\ve{\Delta r_{n}})},
\end{equation}

In our implementation, however, we performed the Fourier transformation before summing over energy bins to speed up the evaluation of different energy bin configurations, taking the special treatment of the diagonal into account. 

The 0.225 mm pixel CdTe system has substantially higher spatial resolution than the other systems (both with 0.5 mm pixels). We therefore also studied the effect of aggregating the pixels in $2\times2$ blocks for the 0.225 mm pixel CdTe system to compare the performance of systems with similar spatial resolution. This aggregation is equivalent to multiplying the transfer function $H^\mathcal{B}_{k,l}(\ve{u})$ with $H_{\mathrm{aggr}}(\ve{u})=\mathrm{cos}(\pi u_x \Delta_x)\mathrm{cos}(\pi u_y \Delta_y)$, where $\Delta_x$ and $\Delta_y$ are the native pixel dimensions, before aggregation. Furthermore, the cross-spectral density is transformed as (cf. Ref. \citenum{cunningham_linear_systems_chapter}, Eq. 2.193)
\begin{equation}
\left[W^{\mathrm{aggr}}_{d^+}(\ve{u})\right]_{k,k'}=\sum_{i=0}^{1} \sum_{i'=0}^{1} \left(H_{\mathrm{aggr}}\left(\ve{u}-\ve{u}_{i,i'}\right)\right)^2 \left[W_{d^+}\left(\ve{u}-\ve{u}_{i,i'}\right)\right]_{k,k'}
\end{equation}

where $\ve{u}_{i,i'}=\left(\frac{i}{2\Delta_x},\frac{i'}{2\Delta_y}\right)^T$ and where the sum over $\mathrm{i}$ and $\mathrm{i'}$ only needs to be taken from 0 to 1 due to the periodicity of $W_{d^+}$.

From the transfer function and cross-spectral density we can now calculate the noise-equivalent quanta (NEQ) matrix:
\begin{equation}
    \label{eq:NEQ_in_basis}
    \mathrm{NEQ}_{l,l'}^{\mathcal{B}}(\ve{u})= \sum_k \sum_{k'} H^\mathcal{B}_{k,l}(\ve{u})^*\left[W_{d^+}(\ve{u})^{-1}\right]^*_{k,k'}\!\! H^\mathcal{B}_{k',l'}(\ve{u}).
\end{equation}

The finite size of the beam used for the point-spread function calculation has the effect of multiplying the transfer function $H^\mathcal{B}_{k,l}(\ve{u})$ with an additional factor $H_{\mathrm{beam}}(\ve{u})=\mathrm{sinc}(u_x b_x)\mathrm{sinc}(u_y b_y)$ where $\mathrm{sinc(x)=\sin(\pi x)/(\pi x)}$. To obtain the true detector NEQ, the calculated NEQ was therefore corrected for this through multiplication with $H_{\mathrm{beam}}(\ve{u})^{-2}$.

We studied system performance both for detection tasks and quantification tasks, as defined in Sec. \ref{sec:introduction}. As our figure of merit for detector performance for \textit{detection} tasks, we use the ``task-specific DQE'' which is the frequency-dependent squared detectability, divided by the squared detectability of an ideal detector with unity detection efficiency and perfect energy and spatial resolution:
\begin{equation}
\label{eq:dqe_task_detection}
\mathrm{DQE}^{\mathrm{Task}}(\ve{u})=\frac{d'(\ve{u})^2}{d_{\mathrm{ideal}}'(\ve{u})^2}=\frac{\ve{\Delta S}^\mathcal{B}(\ve{u})^\dagger \ma{NEQ}^{\mathcal{B}}(\ve{u}) \ve{\Delta S}^\mathcal{B}(\ve{u})}{\ve{\Delta S}^\mathcal{B} (\ve{u})^\dagger \ma{NEQ}^{\mathcal{B},\mathrm{ideal}}(\ve{u}) \ve{\Delta S}^\mathcal{B}(\ve{u})}.
\end{equation}
Here, the relative signal difference $\ve{\Delta S}^\mathcal{B}(\ve{u})$ is $\Delta S_l^{\mathcal{B}}(\ve{u})=\frac{1}{\overline{q}^\mathrm{tot}}\frac{\partial \overline{q}^{\mathrm{tot}}}{\partial A_l}\Delta \tilde{A}_l(\ve{u})$, with the Fourier transformed path length difference $\Delta \tilde{A}_l$ equal to a constant value of $1 \; \mathrm{mm}^{-3}$ for one of the basis materials and zero for the others. The ideal-detector NEQ, $\ma{NEQ}^{\mathcal{B},\mathrm{ideal}}(\ve{u})$, is calculated from Eqs. \ref{transfer_matrix}, \ref{eq:csd} and \ref{eq:NEQ_in_basis}, with $H^{\mathrm{ideal}}_{m}(\ve{u})=1$ and $K^{s,\mathrm{ideal}}_{\ve{\Delta n},k,k'}=\overline{q}_k\delta_{k,k'}$ where $\overline{q}_k$ is the number of counts in energy bin $k$. The energy bins of the ideal detector are assumed to have perfectly sharp borders and contain one energy discretization point each, i.e., $\Delta E=1\; \mathrm{keV}$.

As our figure of merit for detector performance for \textit{quantification} tasks, we use the inverse frequency-dependent CRLB for the noise level in a decomposed basis image $\hat{\tilde{A}}$, normalized by the CRLB obtained from an ideal detector:
\begin{equation}
  \label{eq:dqe_task_quantification}
  \mathrm{DQE}^{\mathrm{Task}}(\ve{u})=\frac{\mathrm{Var}(\hat{\tilde{A}}_l^\mathrm{ideal})}{\mathrm{Var}(\hat{\tilde{A}}_l(\ve{u}))},
\end{equation}

where the variances are proportional to the diagonal elements of the inverse Fisher matrix at frequency $\ve{u}$ (Ref. \citenum{persson_framework_performance_characterization}, eq. 25): $\mathrm{Var}\left(\hat{\tilde{A}}_l(\ve{u})\right) \propto \left[ \mathcal{F(\ve{u})}^{-1}\right]_{l,l}$ with
\begin{equation}
\label{eq:fisher}
\left[\mathcal{F}(\ve{u})\right]_{l,l'}=\frac{1}{{\overline{q}^\mathrm{tot}}^2} \frac{\partial\overline{q}^{\mathrm{tot}}}{\partial A_l}\frac{\partial\overline{q}^{\mathrm{tot}}}{\partial A_{l'}}\mathrm{NEQ}_{l,l'}^{\mathcal{B}}(\ve{u}),
\end{equation}
and correspondingly for $\mathrm{Var}(\hat{\tilde{A}}_l^\mathrm{ideal})$.

The energy bins were optimized separately for each imaging task using the MATLAB function \texttt{fmincon}, with Eqs. \ref{eq:dqe_task_detection} and \ref{eq:dqe_task_quantification} as target functions for detection and quantification tasks, respectively. In both cases, the bins were optimized for $\ve{u}=\ve{0} \; \mathrm{mm}^{-1}$, i.e. for large-area tasks. To avoid converging to a local optimum, 20 randomly drawn threshold configurations were used as starting points. The thresholds were constrained to lie between a minimum threshold $T_\mathrm{min}$ (5 keV for Si, to capture Compton interactions\cite{bornefalk_silicon_strip_pmb}, and 20 keV for CdTe\cite{yu_mayo_research_prototype_evaluation}) and the kVp.

\subsection{Breakdown of DQE}
\label{sec:method_breakdown}
To better understand the detector properties that determine the task-specific DQE for detection, we can express it as a product of factors: 
\begin{equation}
\mathrm{DQE}^{\mathrm{Task}}(\ve{u})=\mathrm{QDE}\cdot f_s\cdot \mathrm{MTF}_{\mathrm{pc}}^2(\ve{u})\cdot f_{\mathrm{mc}}(\ve{u})\cdot \mathrm{DQE}^{\mathrm{Task}}_{\mathrm{ideal \; pc}}\cdot f_{\mathrm{ew}}(\ve{u})
\end{equation}

The quantities in this expression are defined as follows: $\mathrm{QDE}=\overline{q}^{\mathrm{reg}}/\overline{q}^{\mathrm{tot}}$ is the quantum detection efficiency, i.e. the ratio of $\overline{q}^\mathrm{reg}$, the number of unique photons per area registered at least once above the lowest threshold anywhere in the detector, to $\overline{q}^\mathrm{tot}$, the total incident on the detector. $f_s=\left(\frac{\partial \overline{q}^{\mathrm{reg}}}{\partial A_l}\right)^2{\overline{q}^{\mathrm{tot}}}^2/\left(\left(\frac{\partial \overline{q}^{\mathrm{tot}}}{\partial A_l}\right)^2{\overline{q}^{\mathrm{reg\,}}}^2\right)$, where $l$ is the material index to detect, is a spectrum factor that corrects QDE for the fact that the relative difference in transmitted photons is dependent on the material composition of the feature. In our implementation, we have made the simplifying assumption that the lowest threshold is a sharp cutoff (not blurred by electronic noise) when calculating $\overline{q}^{\mathrm{reg}}$ and its derivative. $\mathrm{MTF_{p. c.}}$ is the modulation transfer function in pure photon-counting mode, computed from all interactions including multiple-counted events: $\mathrm{MTF_{p. c.}} = \sum_k H_{k,l}^\mathcal{B}(\ve{u})/\left(\sum_k H_{k,l}^\mathcal{B}(\ve{0})\right)$. $f_{\mathrm{mc}}(\ve{u})=\mathrm{DQE}^{\mathrm{Task}}_{\mathrm{pc}} (\ve{u})/ (\mathrm{DQE}^{\mathrm{Task}}_{\mathrm{ideal \; pc}} \cdot\mathrm{QDE}\cdot f_s\cdot \mathrm{MTF_{pc}}^2(\ve{u}))$ is the factor describing the impact of multiple counting of photons on the $\mathrm{DQE}^{\mathrm{Task}}$ in pure photon-counting mode. $\mathrm{DQE}^{\mathrm{Task}}_{\mathrm{pc}}(\ve{u})$ in the latter equation is the task-specific $\mathrm{DQE}$ in pure photon-counting mode, calculated with the same lowest threshold as for the energy weighting case but only one energy bin, and $\mathrm{DQE}^{\mathrm{Task}}_{\mathrm{ideal \; pc}}$ is the $\mathrm{DQE}^{\mathrm{Task}}$ of an ideal detector in pure photon-counting mode, i.e. a detector which counts all photons in the same energy bin but is ideal in all other respects. Finally, $f_{\mathrm{ew}}(\ve{u})=\mathrm{DQE}^{\mathrm{Task}}(\ve{u})/\mathrm{DQE}^{\mathrm{Task}}_{\mathrm{pc}}$ is the $\mathrm{DQE}$ improvement caused by energy weighting.
\subsection{Simulation of object scatter}
In the simulations described so far, only the primary beam was considered, i.e. scatter from the patient was ignored. To understand the impact of object scatter on the results, we simulated a full CT detector array with an anti-scatter grid in a geometry with a beam diverging from a point source and penetrating a water cylinder, 30 cm in diameter and 30 cm high, at isocenter. The source-to-isocenter and source-to-detector distances were 595 and 1086 mm, respectively. A bowtie filter was included, and had the attenuation profile of the large GE VCT filter in the CatSim simulation suite\cite{catsim}.

The detector was taken as a segment of a sphere centered at the x-ray focal spot. The illuminated area was 79 mm $\times$ 500 mm measured at the isocenter plane, corresponding to the extent of the detector (144 $\times$ 864 mm along the detector arc). To measure the scatter behind the central part of the phantom, however, only interactions taking place in the central 200 mm (measured at the detector) of the detector array were recorded. 

As before, 1.6 mm and 3 mm thick CdTe detectors and 61 mm thick Si detectors with and without internal W foils were simulated. In contrast to the previous simulation, however, all components were taken to be wedge-shaped, enabling the detector to point back to the source without gaps. This geometry leads to a slow variation in pixel pitch across the detector array, expected to have negligible effect on the result. The anti-scatter grid consisted of 25 mm high W lamellae of 100 or 200 $\upmu$m thickness, parallel to the longitudinal axis of the scanner with 1, 2 or 4 mm spacing. A simulation without anti-scatter grid was also made. For the 1.6 mm CdTe design, the dimensions were adjusted to be integer or half-integer multiples of the pixel pitch, i.e. 112.5 or 225 $\upmu$m thickness with 1.125, 2.250 or 4.50 mm spacing. For the Si design with internal W foils, these foils were orthogonal to the longitudinal scanner axis, thereby together with the external anti-scatter lamellae resembling a two-dimensional grid (although the two grids are located at different distances from the source) (Fig. \ref{fig:detector_drawing}(c)). We also simulated two-dimensional anti-scatter grids using two orthogonal one-dimensional grids, both at the same distance in front of the detector. The new lamellae were orthogonal to the longitudinal scanner axis, 25 mm high, and 100 $\upmu$m thick with 1 mm spacing (112.5 $\upmu$m  and 1.125 mm for the 1.6 mm CdTe design) (Fig. \ref{fig:detector_drawing}(d)). This grid design is comparable to one reported in the literature\cite{vogtmeier_anti_scatter_grid}.

For each energy between 20 and 150 keV in steps of 2.5 keV, $3\cdot10^6$ primary photons were simulated, prior to the bowtie filter. All interactions depositing more energy than the lowest threshold (20 keV for CdTe and 5 keV for Si) were registered, and a photon was classified as primary (unscattered) if its direction of incidence pointed at the source and it had the same energy as the primary beam. The total primary and scattered photons for a broad spectrum 120 kVp beam were calculated by weighting the number of primary and scattered photons at each energy with the spectral density of the pre-patient x-ray spectrum described in Sec. \ref{sec:dqe_calculation}, and the total scatter-to-primary ratio, SPR, was obtained by dividing these quantities. 

A DQE modification factor was calculated as $\mathrm{GDE}/(1+\mathrm{SPR})$ (Ref. \citenum{aichinger_x_ray_diagnostic_radiology}, ch. 7) where GDE is the geometric detection efficiency. Note that this approximation ignores the fact that the energy spectrum of the primary and scattered photons are slightly different, which could lead to scatter having different impact on tasks with different spectral dependence. Also, calculating SPR as the ratio of incident unique photons ignores the fact that there could be a slight difference in the probability of double-counting photons for scattered and primary events.

\subsection{Dual-spectrum simulation}
Dual-spectrum imaging with photon-counting detectors was simulated by separately measuring  with a 80 kVp spectrum and a 140 kVp spectrum filtered with 0.4 mm Sn. Energy bin thresholds were optimized as described in Sec. \ref{sec:method_acf}, but with twice the number of degrees of freedom since the thresholds for both spectra were optimized jointly. The zero-frequency detectability obtainable by combining the two measurements was used as the target function for the optimization. To isolate the effect of the combined spectrum shape from the additional energy information from the dual spectra, we also simulated a system that measures the combined 80 and Sn140 kVp spectral densities. This system was simulated with two different energy bin configurations: the configuration found to be optimal for the case with different spectra, and one optimized specifically for the sum-spectrum system, with the thresholds constrained to lie between $T_\mathrm{min}$ and 140 keV. 


In order to compare 80/Sn140 kVp imaging with single-energy imaging at 120 kVp at equal dose, we modeled the $\mathrm{CTDI}_{\mathrm{w}}$ of a commercial dual energy system (Siemens Somatom Flash, Siemens Healthineers, Erlangen, Germany), for a 32 cm CTDI phantom with wide shaped filter and 38.4 mm collimation: 0.011, 0.036 and 0.016 mGy/mAs for 80, 120 and Sn140 kVp, respectively, according to the scanner documentation. Assuming a high kV/low kV mAs ratio of 0.7 \cite{primak_tin_filtration}, this gives equal $\mathrm{CTDI}_{\mathrm{w}}$ for the combination 80/Sn140 kVp as for 120 kVp for the tube current-time product (mAs) of $\left(I\cdot t\right)_{80\;\mathrm{kVp}}=1.6\left(I\cdot t\right)_{120\;\mathrm{kVp}}$ and $\left(I\cdot t\right)_{\mathrm{Sn}140\;\mathrm{kVp}}=1.1\left(I\cdot t\right)_{120\;\mathrm{kVp}}$. Using the air kerma per photon for each of the three beam qualities, and approximating the air kerma as the $\mathrm{CTDI}_{100,\mathrm{air}}$ of the scanner (0.030, 0.090, and 0.039 mGy/mAs for 80, 120 and Sn140 kVp  according to the scanner documentation, for the filter and collimation given above), we obtain the relative photon fluences (photons/$\mathrm{mm}^2$) $\Phi_{80\;\mathrm{kVp}}/\Phi_{120\;\mathrm{kVp}}=0.51$ and $\Phi_{\mathrm{Sn}140\;\mathrm{kVp}}/\Phi_{120\;\mathrm{kVp}}=0.48$. The total number of photons is therefore nearly the same in both the dual energy and single energy cases.

For each spectrum and system, we measured detection performance relative to an ideal, energy-resolving detector with a 120 kVp spectrum. Since the two energy measurements are independent, the relative dose efficiency in the dual-spectrum case is $\frac{d'(\ve{u})^2}{d_{\mathrm{ideal}}(\ve{u})'^2}=\frac{d_L'(\ve{u})^2+d_H'(\ve{u})^2}{d_{\mathrm{ideal,\;120\;kVp}}'(\ve{u})^2}$ where $d_L'(\ve{u})$ and $d_H'(\ve{u})$ are the detectabilities for the low and high kVp measurements, respectively, because the squared detectability (Eq. 6 in Ref. \citenum{persson_framework_performance_characterization}) for the joint measurement is a sum over the squared detectabilities of the two energy measurements. Here, $d_L'(\ve{u})$ and $d_H'(\ve{u})$ are given by
\begin{equation}
d_{L,H}'(\ve{u})^2=\ve{\Delta S}_{L,H}^\mathcal{B}(\ve{u})^\dagger \ma{NEQ}_{L,H}^{\mathcal{B}}(\ve{u}) \ve{\Delta S}_{L,H}^\mathcal{B}(\ve{u})
\end{equation}

We assume that the two energy measurements are perfectly registered so that $\Delta \tilde{A}$ is the same for both tube voltages. Note, however, that $\left[\Delta S_{H,L}^{\mathcal{B}}\right]_l(\ve{u})=\frac{1}{\overline{q}_{H,L}^\mathrm{tot}}\frac{\partial \overline{q}_{H,L}^{\mathrm{tot}}}{\partial A_l}\Delta \tilde{A}_l(\ve{u})$ is different for the low and high kVp, so the NEQ matrices cannot simply be added. Also note that the relative dose efficiency $\frac{d'(\ve{u})^2}{d_{\mathrm{ideal}}'(\ve{u})^2}$ is analogous to the the task-specific DQE, but more general since it captures the effect of the spectrum shape difference as well as the detector performance.


The relative dose efficiency for quantification tasks was calculated as $\frac{\mathrm{Var}(\hat{\tilde{A}}_l^\mathrm{ideal,\;120\;kVp})}{\mathrm{Var}(\hat{\tilde{A}}_l(\ve{u}))}$ with $\mathrm{Var}\left(\hat{\tilde{A}}_l(\ve{u})\right)=\left[ \left(\mathcal{F}_L(\ve{u})+\mathcal{F}_H(\ve{u})\right)^{-1}\right]_{l,l}$ since Fisher matrices are additive for independent sets of measurements as can be shown from Eq. 24 of Ref. \citenum{persson_framework_performance_characterization}. Here, $\mathcal{F}_L(\ve{u})$ and $\mathcal{F}_H(\ve{u})$ are Fisher matrices for the low and high energy measurements given by Eq. \ref{eq:fisher}. 
\section{Results}
\label{sec:results}
\begin{figure}
  \centering
  \includegraphics[scale=0.6]{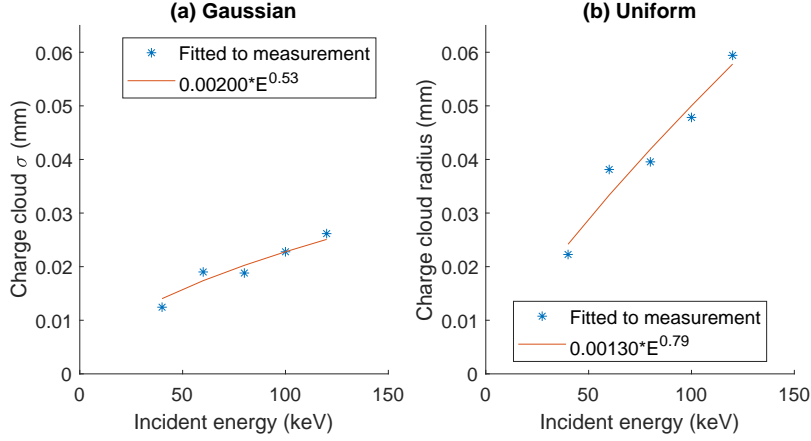}
  \caption[Charge cloud model fitting to silicon energy response.]{Charge cloud model fitting to silicon energy response. (a) Gaussian charge cloud. (b) Uniform charge cloud. Each data point is fitted to a deposited spectrum at a given incident energy, and the power law was fitted to the set of fitted values at different incident energies.}
  \label{fig:chargeModelsSi}
\end{figure}

\begin{figure}
  \centering
  \includegraphics[scale=0.6]{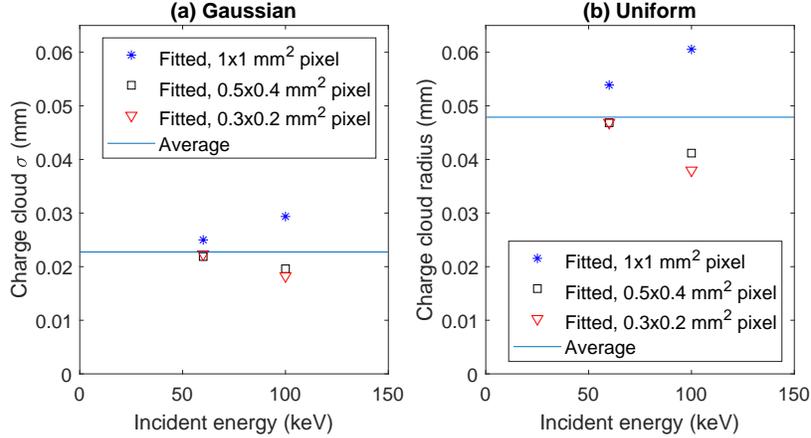}
  \caption[Charge cloud model fitting to CdTe energy response.]{Charge cloud model fitting to CdTe energy response. (a) Gaussian charge cloud. (b) Uniform charge cloud. Uniform charge cloud. Each data point is fitted to a deposited spectrum for one detector pixel size, and one incident energy (60 or 100 keV). The power law was fitted to the set of fitted values at different incident energies.}
  \label{fig:chargeModelsCdTe}
\end{figure}

The Gaussian and uniform charge cloud models fitted to the measured data are shown in Figs. \ref{fig:chargeModelsSi} (for Si) and \ref{fig:chargeModelsCdTe} (for CdTe). For Si, power laws fitted to the charge cloud radii as functions of energy yielded $\sigma=2.0\cdot (E/\mathrm{keV})^{0.53} \;\upmu \mathrm{m}$ for the Gaussian cloud and $r=1.3\cdot (E/\mathrm{keV})^{0.79} \;\upmu \mathrm{m}$ for the uniform cloud. This gives charge cloud radii of $\sigma=17.4 \; \upmu \mathrm{m}$ and $r=33.4 \; \upmu \mathrm{m}$ at 60 keV. For CdTe, no strong energy dependence was observed, so the fitted model was taken to be the average charge cloud size, independent of energy. This average is $\sigma=22.7\; {\upmu}$m for the Gaussian model and $r=47.9\; {\upmu}$m for the uniform model.

\begin{figure}
  \centering
  \includegraphics[scale=0.55]{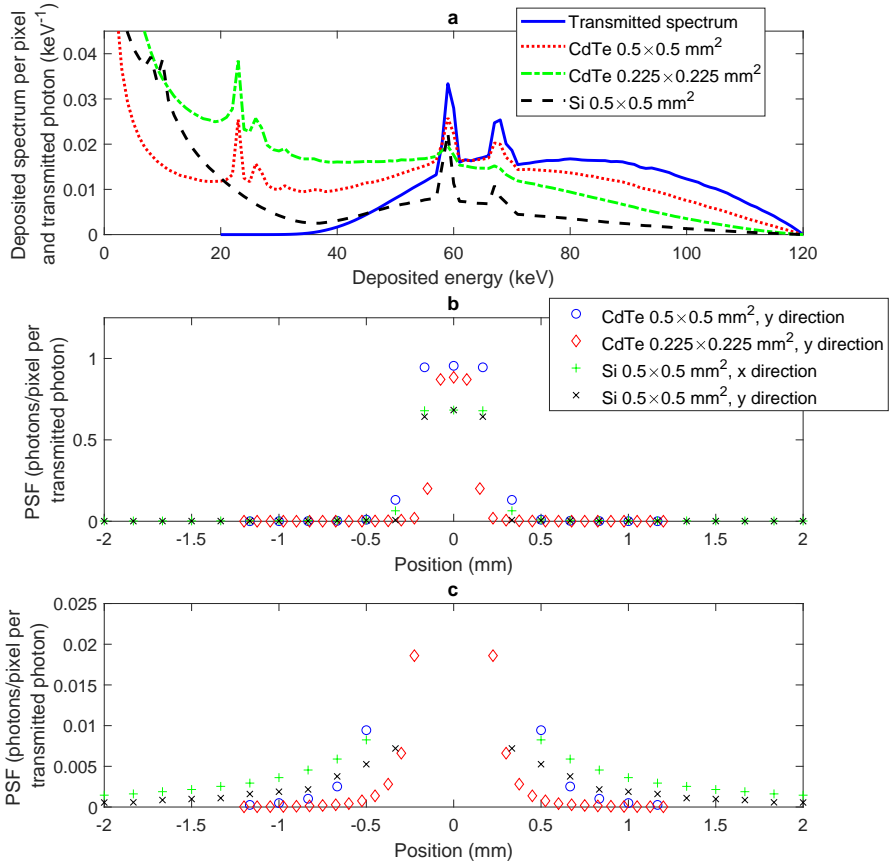} 
  \caption[Deposited energy spectra and point-spread functions.]{(a) Deposited energy spectra for a 120 kVp spectrum attenuated by 300 mm water. (b) Point-spread functions, slices along the x and y axes. (c) Same as (b) with zoomed-in y scale. These quantities are plotted for the CdTe detector designs with $0.5\times0.5 \; \mathrm{mm}^2$ and $0.225\times0.225 \; \mathrm{mm}^2$ pixels with the large charge cloud model, and for the Si detector design with 60 mm active Si and tungsten foils orthogonal to the y axis.}
  \label{fig:PSF_and_spectrum}
\end{figure}

Fig. \ref{fig:PSF_and_spectrum} shows the deposited energy spectra and point-spread functions weighted with the transmitted spectrum $\overline{q}(E)$: 
\begin{equation}
\mathrm{psf^{weighted}}(\ve{r})=\Delta_x \Delta_y \sum_{m:\;E_m>=T_{\mathrm{min}}} \int_E h_m(\ve{r},E)\frac{\overline{q}(E)}{\overline{q}^{\mathrm{tot}}}\mathrm{d}E.
\end{equation}
The sum over $m$ is taken over all deposited energy discretization points greater or equal than the lowest threshold $T_{\mathrm{min}}$ (20 keV for CdTe and 5 keV for Si). Note that the plotted point-spread functions are the intrinsic detector point-spread functions convolved with a square function of side $\frac{1}{3}$ pixel width, representing the aperture of the sub-beam that was used for the Monte-Carlo simulation. Due to the symmetry of the CdTe detector, we only show its point-spread function along the x direction.

\begin{figure*}
  \centering
  \includegraphics[scale=0.65]{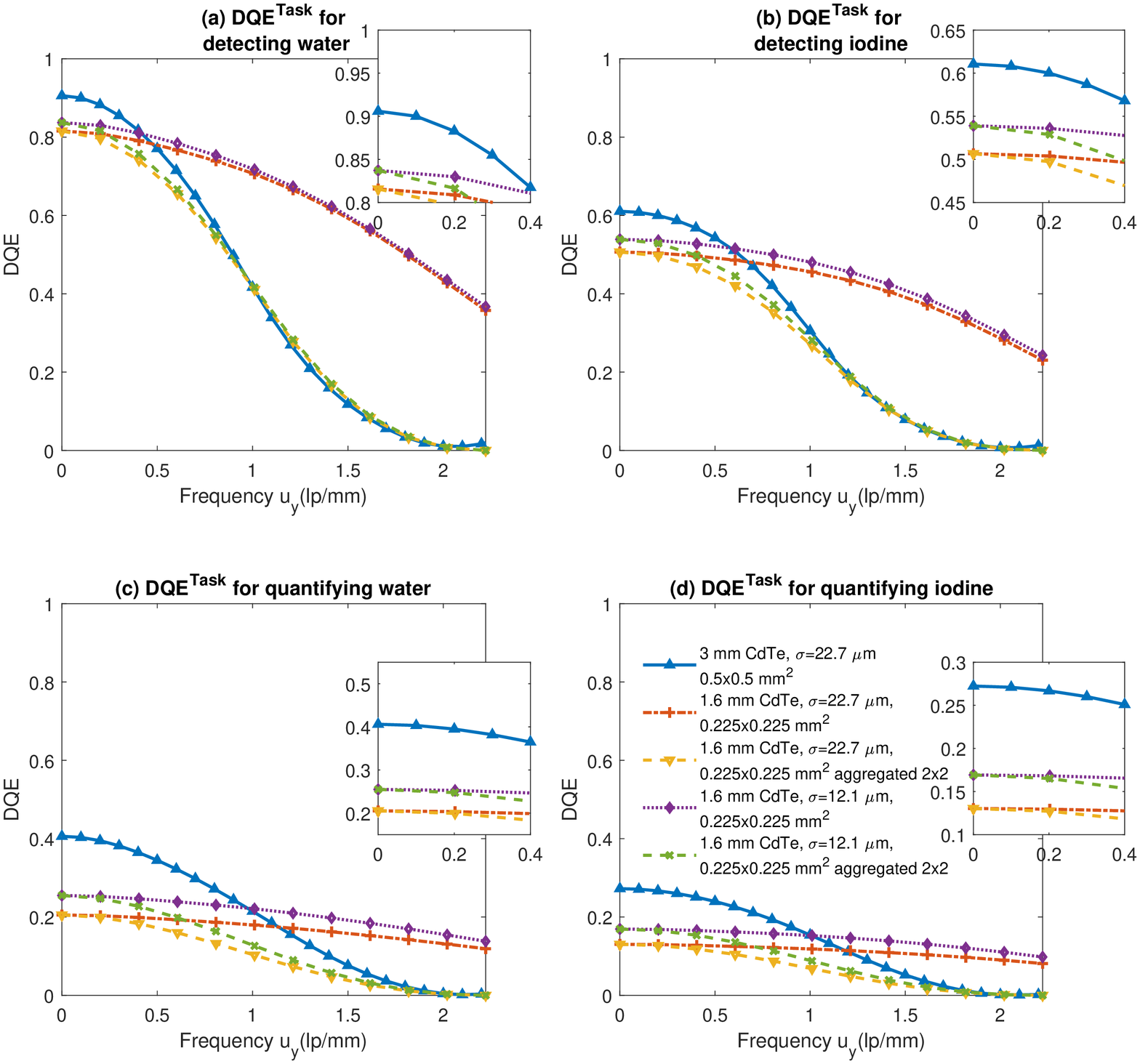}
  \caption[$\mathrm{DQE}^{\mathrm{Task}}$ for CdTe detector designs.]{$\mathrm{DQE}^{\mathrm{Task}}$ along the $u_y$ axis for CdTe detector designs, of 3 and 1.6 mm absorption length. The detector pixel sizes are $0.5 \times 0.5 \; \mathrm{mm}^2$ and $0.225 \times 0.225 \; \mathrm{mm}^2$, both native and aggregated $2\times2$ to $0.45\times0.45\; \mathrm{mm}^2$.  $\sigma$ is the radius of the charge cloud.}
  \label{fig:DQECurvesCdTe}
\end{figure*}

\begin{figure*}
  \centering
  \includegraphics[scale=0.70]{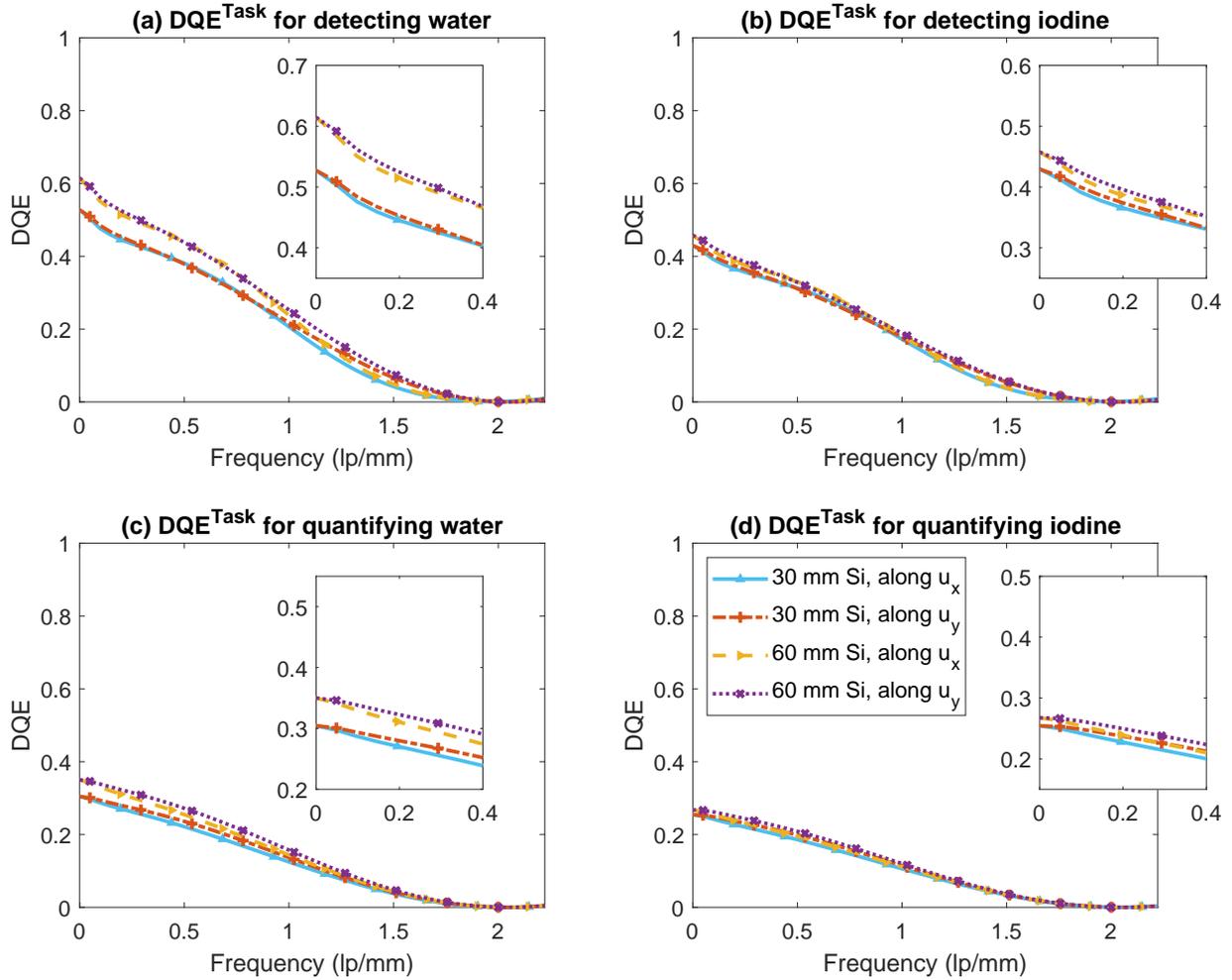}
  \caption[$\mathrm{DQE}^{\mathrm{Task}}$ for Si detector designs with tungsten foils.]{$\mathrm{DQE}^{\mathrm{Task}}$ along the $u_x$ and $u_y$ axes for Si detector designs with tungsten foils between the Si wafers. Two active Si absorption lengths, 30 and 60 mm, are included. The Si wafers and W foils are orthogonal to the y axis.}
  \label{fig:DQECurvesSi_W}
\end{figure*}

\begin{figure*}
  \centering
  \includegraphics[scale=0.70]{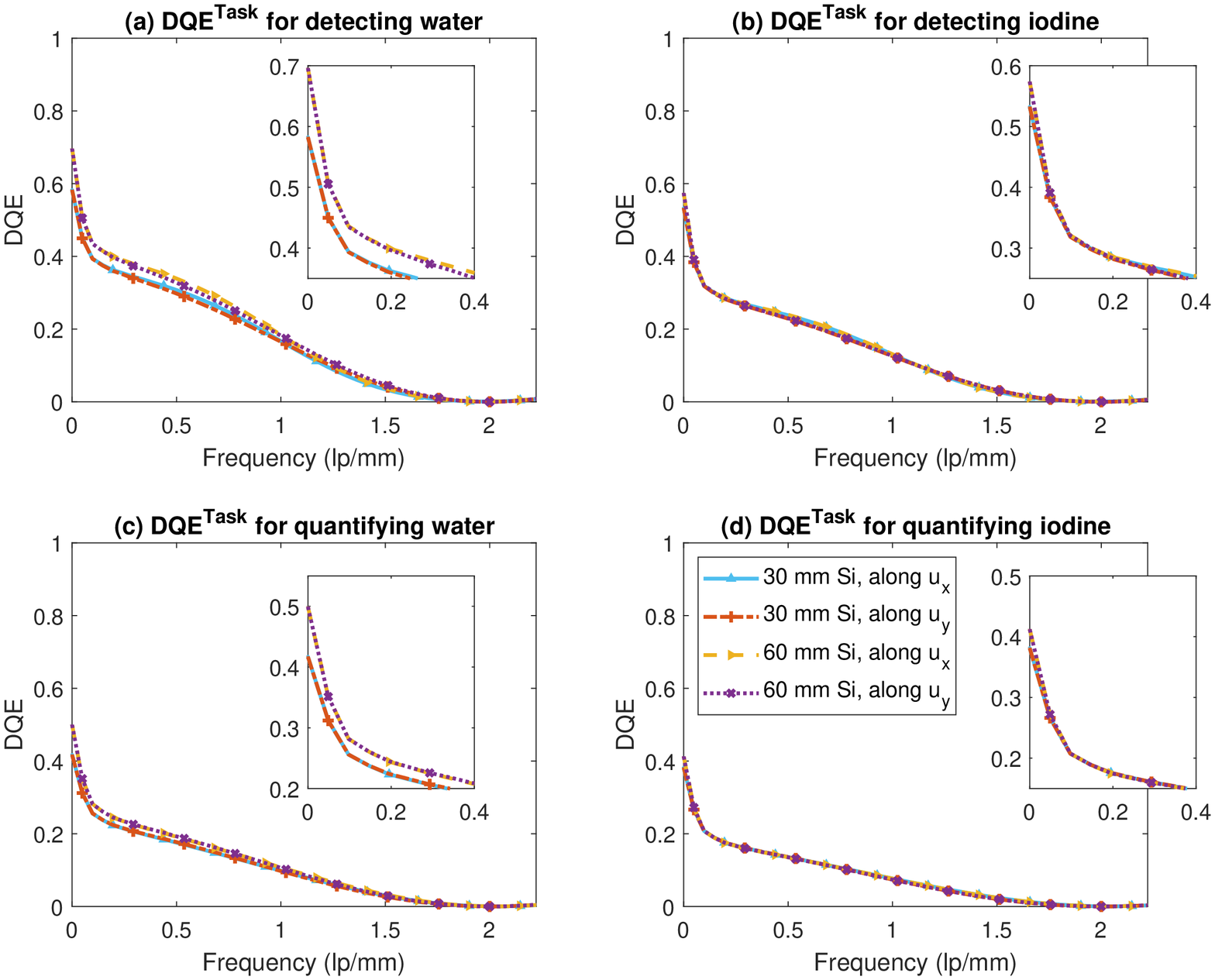}
  \caption[$\mathrm{DQE}^{\mathrm{Task}}$ for Si detector designs without tungsten foils.]{$\mathrm{DQE}^{\mathrm{Task}}$ along the $u_x$ and $u_y$ axes for Si detector designs without tungsten foils. Two active Si absorption lengths, 30 and 60 mm, are included. The Si wafers are orthogonal to the y axis.}  
  \label{fig:DQECurvesSi_noW}
\end{figure*}

Fig. \ref{fig:DQECurvesCdTe} shows the task-specific DQE for the CdTe detectors, for a 120 kVp spectrum and a 300 mm water object. For the water and iodine detection tasks, the energy bins were optimized for these respective tasks, at zero frequency. Since water and iodine are quantified jointly in a two-material decomposition, the plots for the water and iodine quantification tasks are based on the same energy bins, optimized for the zero-frequency DQE for iodine quantification. Figs. \ref{fig:DQECurvesSi_W} and \ref{fig:DQECurvesSi_noW} show the task-specific DQE for a 120 kVp spectrum and a 300 mm water object for the Si detector designs, with and without tungsten foils, respectively. As with the CdTe detector, the energy bins for detection were optimized for their respective tasks at zero frequency, while the energy bins for quantification are optimized for iodine.

Table \ref{table:DQE0} summarizes the zero-frequency DQE for the different detector designs. Also listed in this table is the geometric detection efficiency, which is reduced for the Si design if tungsten (W) foils are included. These numbers build on the idealized assumption of a parallel incident x-ray beam.

\begin{table*}
  \begin{tabular}{lccccc}
  \hline
  \hline
  \shortstack{Detector \\ design}& GDE & \shortstack{Water \\ detection} & \shortstack{Iodine\\ detection} & \shortstack{Water \\ quantification} & \shortstack{Iodine \\ quantification}\\
  \hline
  30 mm Si with W 										& 0.96 & 0.53   & 0.43 & 0.30 & 0.25 \\
  30 mm Si, no W  										& 1    & 0.58   & 0.53 & 0.42 & 0.38 \\
  60 mm Si, W	  										& 0.96 & 0.61	& 0.46 & 0.35 & 0.27 \\
  60 mm Si, no W  										& 1    & 0.70	& 0.57 & 0.50 & 0.41 \\
  \shortstack[l]{3 mm CdTe,\\ 500 ${\upmu}$m pixels}						& 1    & 0.91	& 0.61 & 0.41 & 0.27 \\
  \shortstack[l]{1.6 mm CdTe, \\225 ${\upmu}$m pixels, \\ $\sigma=22.7 \; \upmu\mathrm{m}$} 	& 1    & 0.82	& 0.51 & 0.21 & 0.13\\
  \shortstack[l]{1.6 mm CdTe, \\225 ${\upmu}$m pixels, \\ $\sigma=12.1 \; \upmu\mathrm{m}$}	& 1    & 0.84	& 0.54 & 0.25 & 0.17\\
  \hline
  \hline
  \end{tabular}
  \caption[Zero-frequency DQE for Si and CdTe detector designs.]{Zero-frequency DQE for a 120 kVp spectrum and 300 mm water attenuation, for Si and CdTe detector designs. Anti-scatter grid and object scatter were not included. $\sigma$: standard deviation of Gaussian charge cloud. GDE: geometric detection efficiency.}
  \label{table:DQE0}
\end{table*}

\begin{figure*}
  \centering
  \includegraphics[scale=0.50]{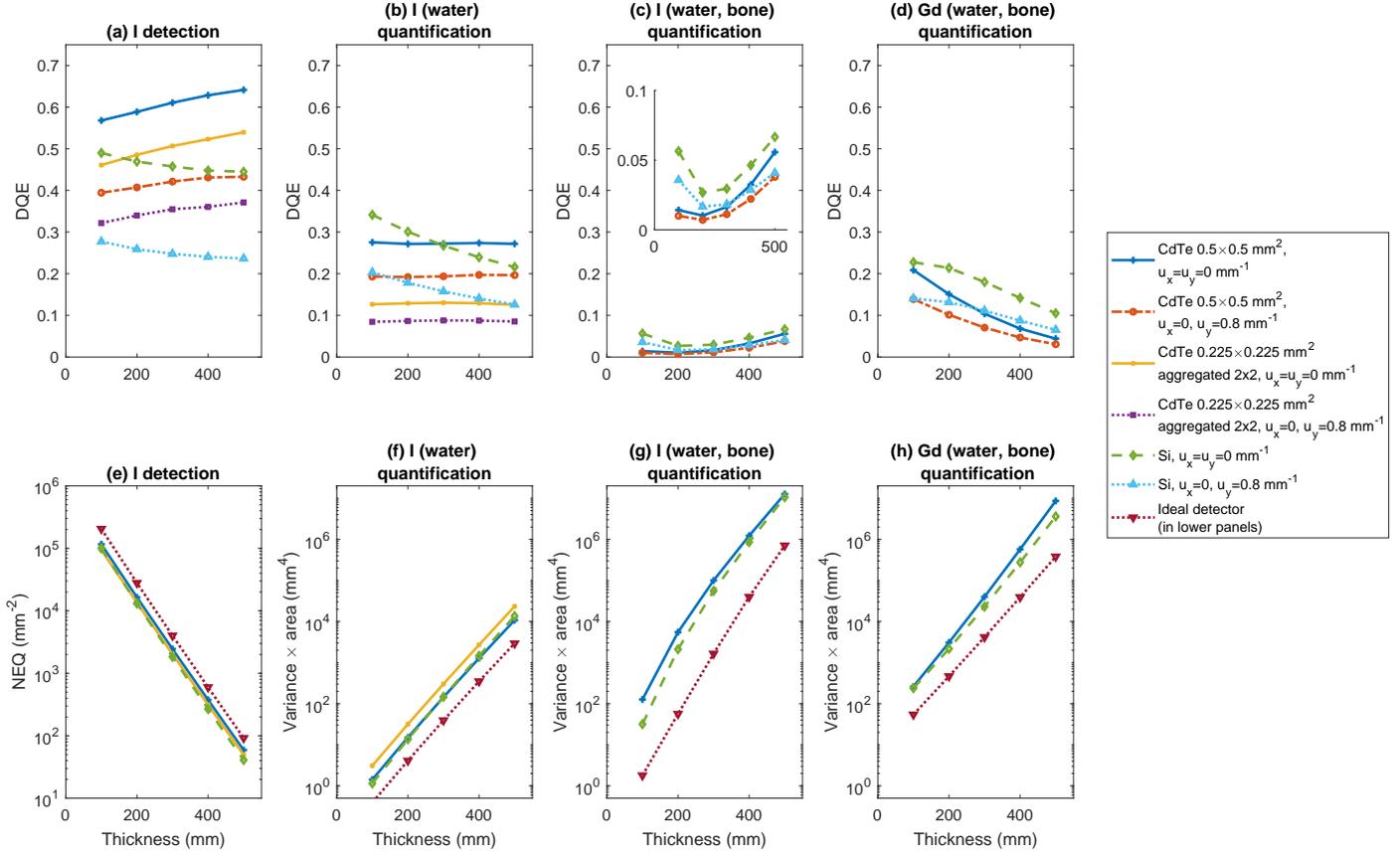}
  \caption[DQE, NEQ and basis image variance vs. attenuating thickness of water.]{DQE, NEQ and basis image variance vs. attenuating thickness of water. The upper panels (a-d) show the DQE at two spatial frequencies: zero frequency and $0.8 \; \mathrm{mm}^{-1}$. The lower panels (e-h) show the NEQ for iodine detection (e) and the I or Gd basis image variance (times measured area) (f-h) at zero frequency, for the studied detectors and the ideal detector. (a, e) Iodine detection. (b, f) Iodine quantification in a two-basis decomposition. (c, g) Iodine quantification in a three-basis decomposition. (d, h) Gadolinium decomposition in a three-basis decomposition. Parentheses denote the materials used in each material decomposition apart from the plotted material. The studied CdTe detectors are the $0.5\times0.5 \; \mathrm{mm}^2$ pixel design and the $0.225\times0.225 \; \mathrm{mm}^2$ pixel design with the large charge cloud model and $2\times2$ pixel aggregation. The Si detector has tungsten foils and 60 mm active absorption length. Three-material decomposition results are not shown for the $0.225\times0.225 \; \mathrm{mm}^2$ design since it has only two energy bins.}
  \label{fig:DQECurvesThickness_Si_andCdTe}
\end{figure*}

Fig. \ref{fig:DQECurvesThickness_Si_andCdTe}(a-d) show the task-specific DQE for the two CdTe and the Si (60 mm active Si, with tungsten foils) detector designs, for 120 kVp and attenuation by a range of water thicknesses. DQEs are plotted for iodine detection, for quantifying iodine in two-and three-material decomposition and for quantifying gadolinium in three-material decomposition. To give an indication about the spatial frequency dependence, the DQE is shown both at zero frequency and at $0.8 \; \mathrm{mm}^{-1}$, i.e. $80\%$ of the Nyquist frequency for detectors with $0.5\times0.5 \; \mathrm{mm}^2$ pixels. The energy bins were optimized for the zero-frequency DQE for each of these tasks. To facilitate the interpretation of these plots, Fig. \ref{fig:DQECurvesThickness_Si_andCdTe}(e-h) show the NEQ and variance for the realistic and ideal detector designs, for a pre-patient fluence of $10^6$ photons/$\mathrm{mm}^2$. The variance is inversely proportional to the total measured area with the plotted proportionality coefficient in $\mathrm{mm}^4$. The DQE (Figs. \ref{fig:DQECurvesThickness_Si_andCdTe}(a-d)) is thus the ratio between the curves for realistic and ideal detectors, realistic/ideal in Fig. \ref{fig:DQECurvesThickness_Si_andCdTe}(e) and ideal/realistic in Fig. \ref{fig:DQECurvesThickness_Si_andCdTe}(f-h).

\begin{figure*}
  \centering
  \includegraphics[scale=0.55]{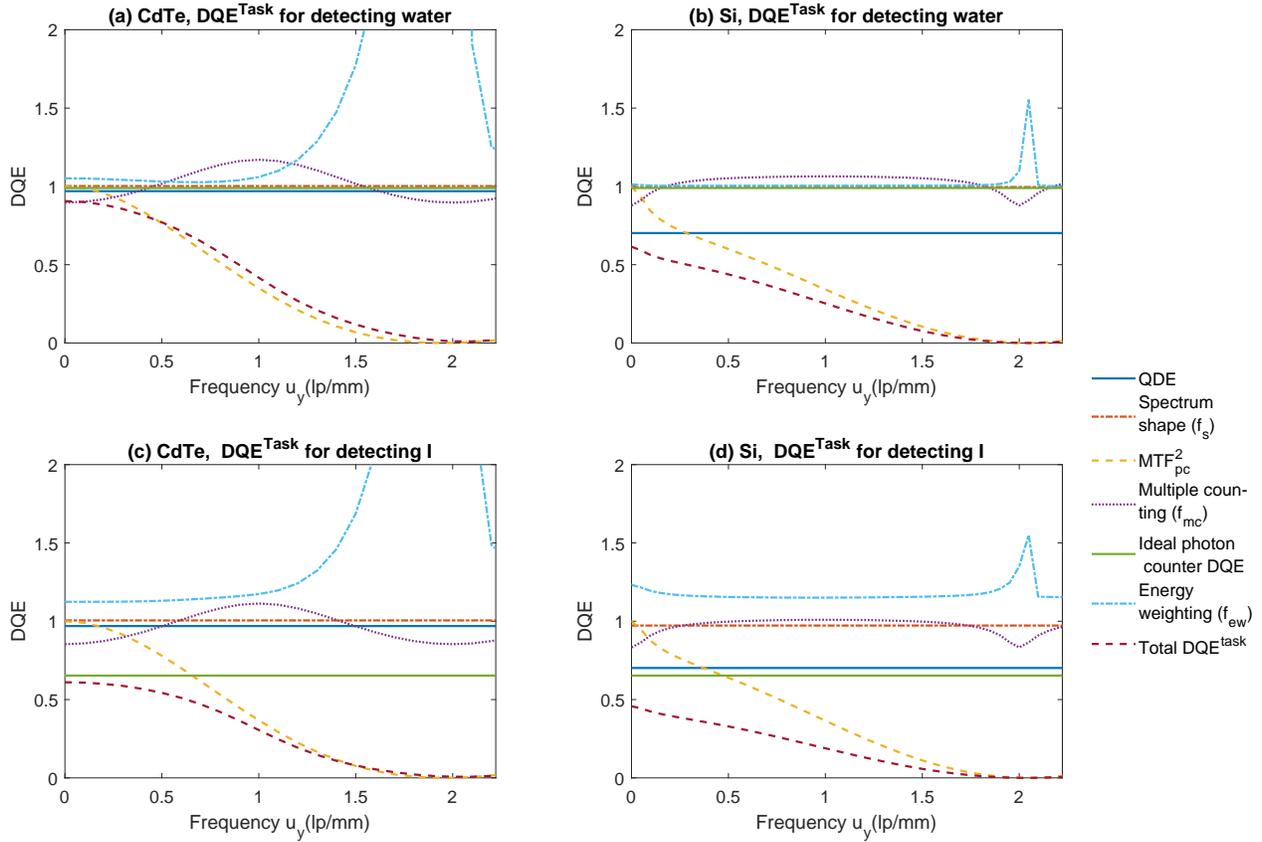}
  \caption[Breakdown of DQE for detection of water and iodine into factors.]{Breakdown of DQE for detection of water and iodine into the factors defined in Sec. \ref{sec:method_breakdown}, plotted along the $u_y$ axis. (a, c): CdTe design with 3 mm absorption depth and $0.5\times0.5 \; \mathrm{mm}^2$ pixels. (b, d): Si design with 60 mm active absorption depth and tungsten foils. The total DQE is the product of the other factors.}
  \label{fig:DQEBreakdown}
\end{figure*}

Fig. \ref{fig:DQEBreakdown} shows the breakdown of the DQE for detection of water and iodine (Figs. \ref{fig:DQECurvesCdTe} and \ref{fig:DQECurvesSi_W}) into a product of factors, as outlined in Sec. \ref{sec:method_breakdown}. The spectrum used here was a 120 kVp spectrum filtered through 300 mm water. Energy bins were optimized for the zero-frequency detection task DQE for water and iodine, respectively.

\begin{figure*}
  \centering
  \includegraphics[scale=0.90]{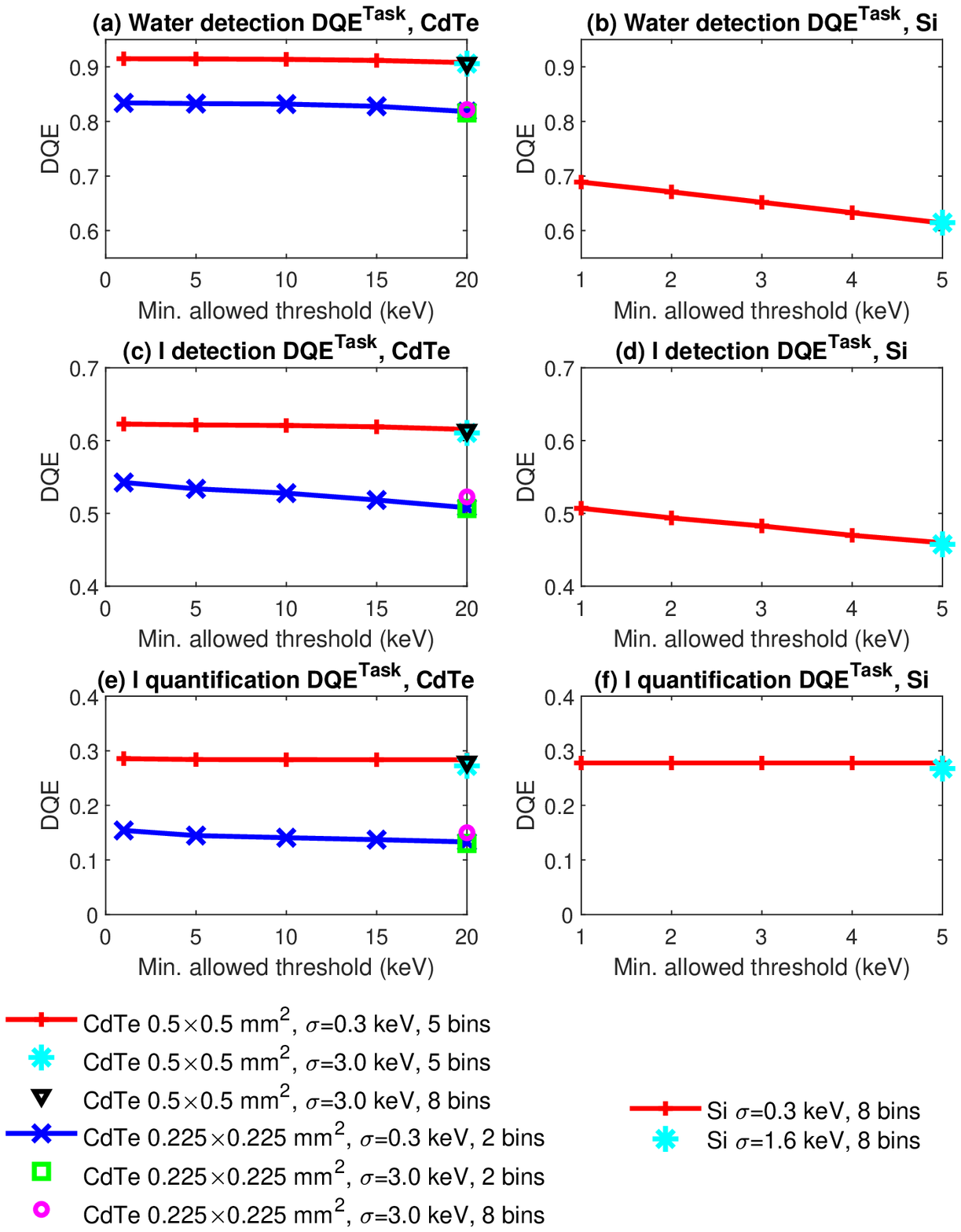}
  \caption[$\mathrm{DQE}^{\mathrm{Task}}$ for CdTe and Si detector designs with lower noise.]{Zero-frequency $\mathrm{DQE}^{\mathrm{Task}}$ for the two CdTe designs (a,c,e) and the 60 mm Si detector design with tungsten foils (b,d,f). For the 1.6 mm thick CdTe detector with $0.225\times 0.225 \;\mathrm{mm}^2$ pixels, the large charge cloud model ($\sigma=22.7 \; \upmu\mathrm{m}$) was used. The $\mathrm{DQE}^{\mathrm{Task}}$ is plotted against the minimum lower threshold permitted in the threshold optimization. Also plotted is the baseline $\mathrm{DQE}^{\mathrm{Task}}$ with the standard electronic noise level (also listed in Table \ref{table:DQE0}), and for CdTe, the corresponding value with eight energy bins.}  
  \label{fig:DQECurvesLowNoise}
\end{figure*}

To understand the potential of each converter material for future improved detectors if the electronic noise could be lowered, the zero-frequency $\mathrm{DQE}^{\mathrm{Task}}$ was also calculated for a range of minimum allowed thresholds from 1 keV to the default minimum threshold of 20 keV (CdTe) and 5 keV (Si) (Figure \ref{fig:DQECurvesLowNoise}). In order for such low thresholds to be realistic without resulting in excessive electronic noise counts, a hypothetical electronic noise level of 0.3 keV RMS was assumed, i.e. substantially lower than the noise in present-day detectors. Also shown is the $\mathrm{DQE}^{\mathrm{Task}}$ for the default values of minimum threshold and electronic noise level (i.e. the value shown in Table \ref{table:DQE0}), and the corresponding values for CdTe with eight energy bins. Note that the parameter plotted on the x axis is the minimum value that the lowest threshold is allowed to assume in the bin optimization algorithm, not the actual lowest threshold. In some cases the actual lowest threshold resulting from the threshold optimization is higher than this, such as for iodine quantification with the Si detector where the lowest threshold is 8.4 keV independent of the threshold constraint.

The scatter-to-primary ratio resulting from the object scatter simulations is shown in Table \ref{table:SPR}. The corresponding scatter DQE factor $\mathrm{GDE}/(1+\mathrm{SPR})$ is shown in table \ref{table:DQE_scatter_factor}. The geometric efficiency (GDE) is the product of two factors: the GDE of the grid itself and the intrinsic GDE of the detector. Both factors are included in Table \ref{table:DQE_scatter_factor}. Note that the value for the Si design with internal W foils and two-dimensional anti-scatter grid was obtained under the assumption that the external grid obscures every second internal foil, giving an intrinsic detector GDE of 0.98. The DQE values resulting from combining the intrinsic zero-frequency DQE of Table \ref{table:DQE0} with this scatter DQE factor are shown in Table \ref{table:DQE0_with_scatter}.

\begin{table*}
  \begin{tabular}{lcccccc}
  \hline
  \hline
  \shortstack{Detector \\ design}& No grid & \shortstack{1D,\\ 0.1 mm W \\ every 4 mm} & 	\shortstack{1D, \\0.1 mm W\\ every 2 mm} & \shortstack{1D, \\0.1 mm W\\ every 1 mm} & \shortstack{1D, \\0.2 mm W\\ every 1 mm} & \shortstack{2D, \\0.1 mm W\\ every 1 mm}\\
  \hline
  60 mm Si, W&	 0.57&	0.21&	0.13&	0.07&	0.06&	0.03\\
  60 mm Si, no W &1.17&	0.43&	0.24&	0.14&	0.13&	0.06 \\
  \shortstack[l]{3 mm CdTe,\\ 500 ${\upmu}$m pixels} & 1.15&	0.40&	0.22&	0.11&	0.10&	0.03\\
  \shortstack[l]{1.6 mm CdTe, \\225 ${\upmu}$m pixels*}& 1.20&	0.46&	0.25&	0.13&	0.12&	0.04\\
  \hline
  \shortstack{GDE of anti-\\scatter grid}& 1&	0.975&	0.95&	0.9&	0.8&	0.81\\
 \hline
  \end{tabular}
  \caption[Monte-Carlo simulated scatter-to-primary ratio (SPR).]{Monte-Carlo simulated scatter-to-primary ratio (SPR) for different detector and anti-scatter grid designs. *For the CdTe detector with 225 ${\upmu}$m pixels, lamellae thicknesses and spacings were $12.5 \%$ larger, to match the pixel pitch.}
  \label{table:SPR}
\end{table*}
\begin{table*}
  \begin{tabular}{lccccccc}
  \hline
  \hline
  \shortstack{Detector \\ design} & \shortstack{Intrinsic\\ GDE} & No grid &\shortstack{1D,\\ 0.1 mm W \\ every 4 mm} & 	\shortstack{1D, \\0.1 mm W\\ every 2 mm} & \shortstack{1D, \\0.1 mm W\\ every 1 mm} & \shortstack{1D, \\0.2 mm W\\ every 1 mm} & \shortstack{2D, \\0.1 mm W\\ every 1 mm}\\
  \hline
  60 mm Si, W&	0.96& 0.61&	0.77&	\textbf{0.81}&	\textbf{0.81}&	0.72&	0.77**\\
  60 mm Si, no W&	1& 0.46&	0.68&	0.76&	\textbf{0.79}&	0.71&	0.77\\
  \shortstack[l]{3 mm CdTe,\\ 500 ${\upmu}$m pixels} &   1 & 0.47&	0.70&	0.78&	\textbf{0.81}&	0.73&	0.78\\
  \shortstack[l]{1.6 mm CdTe, \\225 ${\upmu}$m pixels*}&   1 & 0.45&	0.67&	0.76&	\textbf{0.80}&	0.71&	0.78\\
  \hline
  \shortstack{GDE of anti-\\scatter grid}& &1   &	0.975&	0.95&	0.9&	0.8&	0.81\\
 \hline
  \end{tabular}
  \caption[DQE factor calculated from the scatter-to-primary ratio.]{DQE factor $\mathrm{GDE}/(1+\mathrm{SPR})$ calculated from the scatter-to-primary ratio in Table \ref{table:SPR} and the geometric detection efficiency (GDE). The GDE is the product of the anti-scatter grid DQE and the intrinsic GDE of the detector, as included in the table. *For the CdTe detector with 225 ${\upmu}$m pixels, lamellae thicknesses and spacings were $12.5 \%$ larger, to match the pixel pitch. **Effective intrinsic GDE for the Si design with W and a 2D grid is 0.98 since the external grid shadows some of the internal W foils.}
  \label{table:DQE_scatter_factor}
\end{table*}

\begin{table*}
  \begin{tabular}{lcccc}
  \hline
  \hline
  \shortstack{Detector \\ design}& \shortstack{Water \\ detection} & \shortstack{Iodine\\ detection} & \shortstack{Water \\ quantification} & \shortstack{Iodine \\ quantification}\\
  \hline
  60 mm Si, W	  										 & 0.52 & 0.39 & 0.29 &	0.23 \\
  60 mm Si, no W  										 & 0.55	& 0.45 & 0.40 &	0.33 \\
  \shortstack[l]{3 mm CdTe,\\ 500 ${\upmu}$m pixels}						 & 0.73 & 0.49 &	0.33 &	0.22\\
  \shortstack[l]{1.6 mm CdTe, \\225 ${\upmu}$m pixels, \\ $\sigma=22.7 \; \upmu\mathrm{m}$} 	& 0.65 & 0.40 & 0.16 & 0.10\\

    \shortstack[l]{1.6 mm CdTe, \\225 ${\upmu}$m pixels, \\ $\sigma=12.1 \; \upmu\mathrm{m}$} 	& 0.67 & 0.43 & 0.20 & 0.13\\
  \hline
  \end{tabular}
  \caption[Scatter-adjusted zero-frequency DQE with object scatter included.]{Zero-frequency DQE for a 120 kVp spectrum and 300 mm water attenuation, for Si and CdTe detector designs, with scatter from the imaged object and anti-scatter grid. The 1D grid of 0.1 mm thick lamellae with 1 mm spacing was used for all detector designs.}
  \label{table:DQE0_with_scatter}
\end{table*}

\begin{figure*}
  \centering
  \includegraphics[scale=0.55]{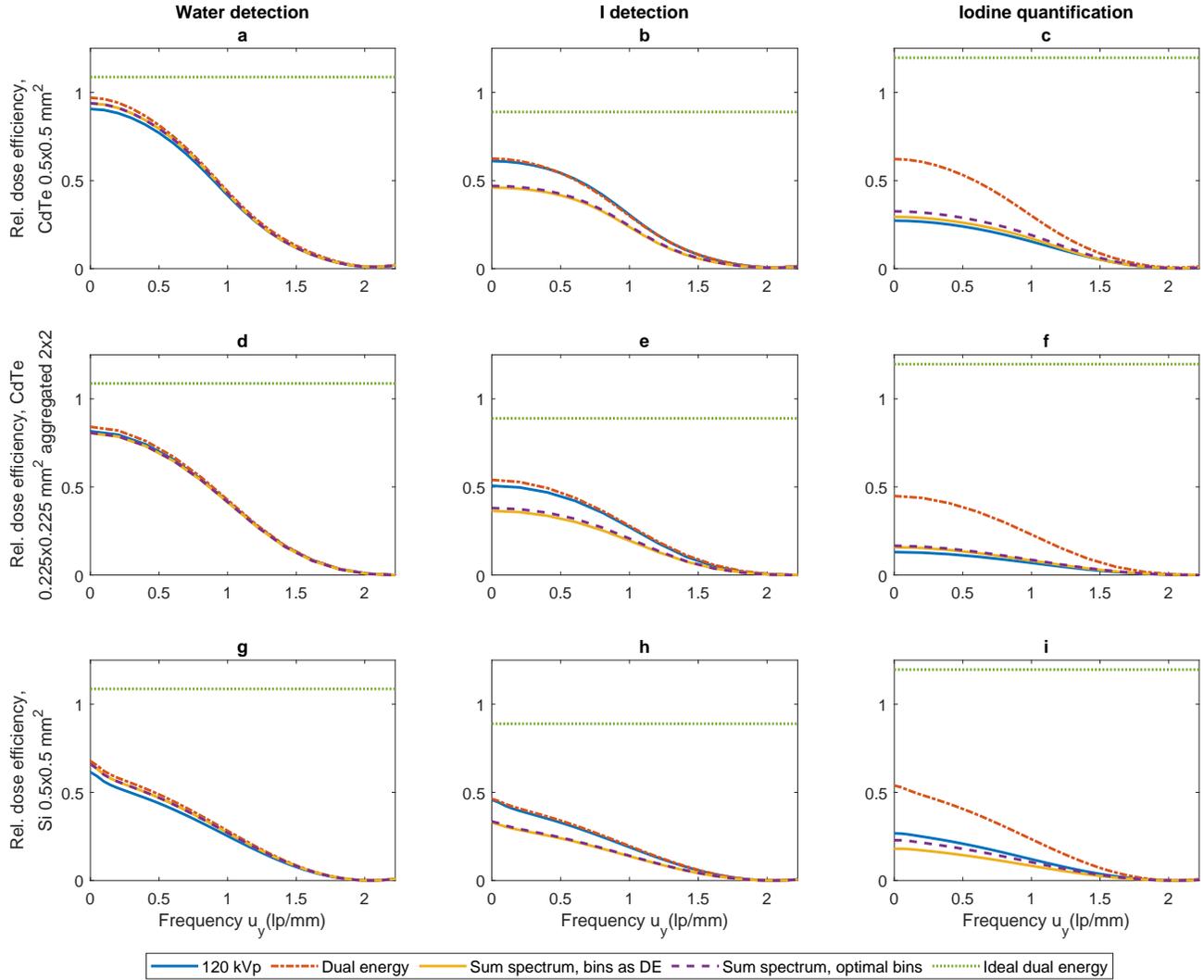}
  \caption[Relative dose efficiency for dual energy imaging relative to single-energy imaging.]{Relative dose efficiency for water and iodine detection and two-basis iodine quantification, for dual energy imaging (80/Sn140 kVp) relative to single-energy (120 kVp) imaging, both with an ideal energy-resolving detector. Also plotted is the 120 kVp single-energy DQE of the studied detectors.``Sum spectrum'': performance when both detectors measure the sum of the 80 and Sn140 kVp spectra. DE bins: same bin configuration as used for dual energy. Optimal bins: bins optimized for the sum spectrum case. ``Ideal dual energy'': relative dose efficiency of 80/140 kVp dual-spectrum imaging with two ideal energy-resolving detectors. 300 mm water attenuation as assumed for all spectra.}
  \label{fig:DQECurvesDE_Si_andCdTe} 
\end{figure*}

Fig. \ref{fig:DQECurvesDE_Si_andCdTe} the shows detectability of 80/Sn140 kVp dual energy imaging relative to 120 kVp imaging with an ideal detector with perfect detection efficiency and spatial and energy resolution. The effects of object scatter and anti-scatter grid are not included in these results. To isolate the effect of the spectrum shape from the effect of dual-spectrum imaging, Fig. \ref{fig:DQECurvesDE_Si_andCdTe} also shows the performance for a dual-spectrum system where both detectors are illuminated by the same spectrum: the sum of the 80 and Sn140 kVp spectra. This is plotted for the CdTe detectors with $0.5\times0.5 \; \mathrm{mm}^2$ and $0.225\times0.225 \; \mathrm{mm}^2$ pixels, the latter with the same charge cloud model as the $0.5\times0.5 \; \mathrm{mm}^2$ design, and for the Si design with 60 mm active Si and tungsten foils. Each of the plotted systems has energy bins optimized for the respective task (water or iodine detection or iodine quantification) at zero frequency. For comparison, the combined-spectrum performance is also shown with the energy bin configuration optimized for dual energy.

\section{Discussion}
\label{sec:discussion}
The fitted charge cloud models in Fig. \ref{fig:chargeModelsSi} and Fig. \ref{fig:chargeModelsCdTe} exhibit different energy dependences. For silicon (Fig. \ref{fig:chargeModelsSi}) the cloud radius increases with increasing incident energy, with power law exponents close to 0.5 and 0.8 for Gaussian and uniform charge clouds. Note, however, that the fitted curve is nearly linear within the studied energy range so the exact values of the exponents are expected to be sensitive to small uncertainties in the data points. For the fitted Gaussian model for Si, the charge cloud standard deviation increases from 14 to 25 $\upmu \mathrm{m}$ between 40 and 120 keV. For CdTe, on the other hand, different fitted charge cloud sizes result from fitting to different pixel sizes, but the fits do not exhibit any clear energy dependence (Fig. \ref{fig:chargeModelsCdTe}). The discrepancy for different pixel sizes may be caused by model discrepancies between the present simulation model and the model of Ref. \citenum{xu_charge_sharing}. For example, incomplete charge collection is modeled in the referenced publication but not in the present model, and this means that the present model has a sharper photopeak than the reference spectrum, in particular for the high energy (100 keV) data point. The RMS charge cloud radii presented here can be compared to a simulated radius of approximately 45 $\upmu \mathrm{m}$ encompassing 95\% of the charge, reported in Ref. \citenum{lai_czt_modeling}.

Note that a difference between the Si and CdTe models is that the Si model is fitted to a measurement while the CdTe model is fitted to a simulation model. The CdTe model may therefore be a lower bound to the performance of a real CdTe detector with less-than-ideal material properties.

When fitting the charge sharing model we made the simplifying assumption that the charge cloud is dependent only on the incident monochromatic energy and independent of the actual deposited energy in each interaction. Despite this, we used the resulting fitted charge cloud sizes to derive a relationship between deposited x-ray energy and charge cloud radius. This simplification ignores the fact that some photons deposit less than their full energy when they interact in the material. However, this approximation is justified, in CdTe because the resulting fitted model for the charge cloud radius is independent of energy, and in Si because the deposited spectrum is dominated by the photopeak (the Compton interactions have too low energies to be included in the fitting and there are few reabsorptions of scattered photons in the single-wafer geometry).

The deposited spectra shown in Fig. \ref{fig:PSF_and_spectrum}(a) reflect the different physics in the different detector materials. Both CdTe detector designs exhibit strong spectral tailing towards low energies, due to charge sharing and fluorescence. This tail is largest for the 0.225 mm detector since the smaller pixel size means that more photons deposit energy in more than one pixel. In silicon, on the other hand, fluorescence does not play a large role and the effect of charge sharing is reduced since it effectively only occurs in one dimension, i.e. within each silicon wafer. Instead, the dominant physical effect degrading the spectrum is Compton scatter, which leads to a peak in the spectrum towards zero keV. Note that there is little overlap between the Compton scatter and the photoelectric interactions, i.e. the photoelectric part of the spectrum is reduced in magnitude due to Compton scatter but its spectral content is largely unaffected.

This difference between Si and CdTe is also reflected in the point-spread functions in Fig. \ref{fig:PSF_and_spectrum}(b) The CdTe point-spread functions are blurred by fluorescence and charge sharing and therefore fall off to zero within a few hundred $\upmu\mathrm{m}$ outside the pixel border. On the other hand, the Si point-spread function exhibits several mm long tails caused by Compton scatter, in particular along the x direction where the scatter is not blocked by tungsten foils. 


The effect of the tungsten foils on intradetector scatter can also be seen by comparing the DQE curves for silicon with and without tungsten foils (Fig. \ref{fig:DQECurvesSi_W} and \ref{fig:DQECurvesSi_noW}), which show that inclusion of tungsten foils decreases the zero-frequency DQE by 9-35 $\%$ (Table \ref{table:DQE0}) while it increases DQE at higher frequencies. Without tungsten foils, the DQE has a sharp peak at low frequencies, corresponding to the long range of the Compton scatter. For detecting an object that is larger than the Compton scatter range, absorption of Compton photons makes a positive contribution to the signal, whereas their only effect when imaging a smaller object is to increase noise. In addition, the tungsten foils decrease the geometric detection efficiency by $4\%$ by replacing some of the silicon. The tungsten foils, therefore, remove some useful signal at very low frequencies, but in most practical situations this drawback is likely outweighed by their positive effect on DQE at higher frequencies.

Apart from decreasing the amount of Compton scatter, the tungsten foils also decrease the pixel aperture in the $y$ direction, thereby leading to a DQE improvement at high frequencies close to the sampling frequency $(2 \; \mathrm{mm}^{-1}$) as seen in Fig. \ref{fig:DQECurvesSi_W}(a-b). Other differences between the x and y directions in Fig. \ref{fig:DQECurvesSi_W} are caused by the fact that charge sharing only takes place along the $x$ (within-wafer) direction and that Compton scatter is more prominent in this direction.

The DQE curves for CdTe (Fig. \ref{fig:DQECurvesCdTe}) show that the system with smaller pixels has a substantial benefit for high frequencies due to its broader MTF. We also studied the effect of aggregating the native pixels in blocks of $2\times2$, similar to the mode described in Ref. \citenum{leng_uhr}. This aggregation reduces the requirements of the data readout system. In our study it allows us to examine the effect of varying the native pixel size while keeping the overall system resolution similar. As seen in Fig. \ref{fig:DQECurvesCdTe}, the $2\times2$ aggregation yields an overall frequency dependence resembling that of the $0.5\times0.5 \; \mathrm{mm}^2$ pixels, but with low-frequency DQE equal to that of the system with native pixels. The degradation of high-frequency DQE caused by the $2\times2$ aggregation corresponds to the known phenomenon that aggregating pixels leads to increased image noise if the resolution is kept fixed.\cite{baek_to_bin_or_not} Relative to the 0.5 mm pixel system, the low-frequency DQE of the 0.225 mm system is 8-10 $\%$ lower for water detection and 12-17 $\%$ lower for iodine detection depending on the charge cloud model. For quantification tasks, the difference is larger: 37-52 $\%$, due to increased charge sharing caused by the smaller pixel size, fewer energy bins and thinner converter of the 0.225 mm design. The increased charge sharing and smaller number of energy bins have a large negative impact on energy resolution, which is why the quantification tasks are affected particularly strongly.

We did not have charge cloud data for a 1.6 mm thick CdTe sensor. We estimate that the charge cloud size will lie between the value fitted to 3 mm CdTe ($\sigma=22.7 \; \upmu\mathrm{m}$) and the value obtained by rescaling this size linearly with the sensor thickness ($\sigma=12.1 \; \upmu\mathrm{m}$). We found that the DQE difference between these two models is small for detection (3-6 $\%$ difference) but larger for quantification (24-30 $\%$) (Table \ref{table:DQE0}).

The plots of detector performance vs. thickness (Fig. \ref{fig:DQECurvesThickness_Si_andCdTe}) reveal the different behavior of the different detector designs when the thickness of the background material (water) is varied. Unlike the CdTe designs, the DQE for the 60 mm Si detector with tungsten foils decreases with increasing patient thickness for iodine detection and two-material decomposition, so that the Si detector outperforms both CdTe detectors for zero frequency at 100 mm thickness but has inferior performance compared to the 0.5 mm CdTe detector for 500 mm thickness. This indicates that the drawbacks of silicon, i.e. limited absorption efficiency and high probability of Compton scatter, become more severe at higher energies and are therefore are more detrimental when imaging highly attenuating objects. 

The three-material quantification DQE (Fig. \ref{fig:DQECurvesThickness_Si_andCdTe}(c-d)) shows different behavior depending on whether the contrast agent is iodine or gadolinium. The DQE for imaging gadolinium decreases with increasing overall object thickness, suggesting that the increased fraction of Compton scatter in Si and spectral tailing due to fluorescence and charge sharing in CdTe make the K-edge of gadolinium harder to detect with increasing hardness of the transmitted x-ray spectrum. When quantifying iodine in a three-basis decomposition, on the other hand, the DQE drops to a minimum at 200-300 mm thickness and then increases again. To understand why, it is useful to study the variance of the simulated and ideal detectors (Fig. \ref{fig:DQECurvesThickness_Si_andCdTe}(g)). While the variance for the simulated and ideal CdTe and Si detectors both increase over the studied range of patient thicknesses, the variance of the simulated detectors increases more rapidly for small thicknesses than for large thicknesses, giving rise to the U-shaped DQE curve. Since the K-edge of iodine is located at the lower end (33.2 keV) of the diagnostic energy range, it can be used to distinguish iodine from other materials if the imaged patient is thin, but very little of the x-ray spectrum near the K-edge remains after filtration through 200-300 mm soft tissue. For thicker patients, the simulated three-material decomposition relies on higher-energy features distinguishing the iodine attenuation coefficient from the other basis functions, and compared to the K-edge, these higher-energy features are less affected by detector imperfections such as Compton scatter and spectrum tailing. Also note that the variance increases more rapidly with increasing thickness in Fig. \ref{fig:DQECurvesThickness_Si_andCdTe}(g) compared to the other imaging tasks (Fig. \ref{fig:DQECurvesThickness_Si_andCdTe}(e, f, h)), indicating that three-basis iodine quantification is more difficult for large patient thicknesses. It remains to be investigated whether the features of the iodine linear attenuation coefficient used by this simulation correspond to real-world physical features or if they come from inaccuracies in tabulated coefficients.

Comparing CdTe and Si for three-basis decomposition, the 0.5 mm pixel CdTe design has lower dose efficiency (42-44 $\%$) compared to the 60 mm Si design with tungsten foils, for 300 mm water attenuation (Fig. \ref{fig:DQECurvesThickness_Si_andCdTe}). The largest difference is found for three-basis iodine decomposition with 100 mm water filtration, where the DQE of the CdTe detector with large pixels is 75 $\%$ lower than the DQE of the 60 mm Si detector. This reflects the trend that silicon, with its limited amount of spectral overlap, performs better relative to CdTe for tasks more heavily dependent on energy resolution. Three-material decomposition is not included for the 0.225 mm pixel detector in figure \ref{fig:DQECurvesThickness_Si_andCdTe} since two energy bins are insufficient for three-material decomposition unless extra constraints, such as volume conservation, are enforced. We would like to point out that a real-world prototype detector\cite{yu_mayo_research_prototype_evaluation} that resembles our model system has a ``chess mode'' where the pixels are divided into two subsets according to a checkerboard pattern and the thresholds in the two subsets are set to two different configurations. By summing the individual pixels into macro-pixels, this effectively gives four energy bins. Simulating such a mode is however out of the scope of this investigation.

Fig. \ref{fig:DQECurvesLowNoise} shows the impact of the lower threshold limit for different detector designs. For the CdTe detector with $0.5\times 0.5 \: \mathrm{mm}^2$ pixels, the DQE is more or less independent on the lowest allowed threshold. On the other hand, the CdTe detector with $0.225\times 0.225 \: \mathrm{mm}^2$ exhibits a DQE improvement from 0.51 to 0.54 for the iodine detection task and from 0.13 to 0.15 for the iodine quantification task, as the threshold constraint is lowered from 20 keV to 1 keV. This suggests that some of the events lost due to charge sharing could improve performance if allowed to contribute. For the Si detector, quantification DQE is unaffected while the detection DQE is improved substantially, from 0.61 to 0.69 for water and from 0.46 to 0.51 for iodine when the threshold constraint is lowered from 5 keV to 1 keV. This illustrates the fact that Compton events are useful for detection tasks since they improve the quantum efficiency, but do not contain much energy information. Another conclusion from Fig. \ref{fig:DQECurvesLowNoise} is that improving the RMS energy resolution due to electronic noise to 0.3 keV has a limited effect on the DQE. However, increasing the number of energy bins from two to eight improves the iodine quantification DQE from 0.13 to 0.15 for the CdTe detector with $0.225\times0.225 \: \mathrm{mm}^2$ pixels and a 20 keV lower threshold, suggesting that using more than two thresholds is desirable for spectral imaging tasks.

The breakdown of the task-specific DQE into its constituent factors (Fig. \ref{fig:DQEBreakdown}) can be helpful to elucidate the factors that influence the DQE. The coarsest estimate of the dose efficiency is the quantum detection efficiency (QDE), i.e. the fraction of incident photons that are registered at least once by the detector. The QDE is 0.97 for 3 mm thick CdTe and 0.70 for 60 mm Si, showing the advantage of the higher stopping power of CdTe. A small adjustment to this estimate is obtained by multiplying it with the spectrum factor $f_s$ between 0.97 and 1.01, which corrects for the fact that different incident energies contain different amounts of information about the object, even when the detector itself is not energy discriminating. The main factor determining the spatial frequency dependence of the result is the squared MTF, which is plotted for purely photon-counting (single-energy-bin) mode. However, the MTF only captures the frequency dependence of the signal, and not that of the correlated noise, which is why the shape of the MTF needs to be corrected by a factor $f_{\mathrm{mc}}$ which accounts for multiple counting. The frequency dependence of this factor show that the low-frequency character of the correlated noise: a concentration of the NPS at low frequencies leads to a degraded DQE at these frequencies. For the CdTe detector, these correlations are caused by fluorescence and charge sharing and are short-ranged (on the order of a pixel size). For the Si detector on the other hand, the correlations are caused by Compton scatter which has longer range, and the effect of multiple counting is therefore concentrated at lower frequencies.

The product of the above four factors gives the dose efficiency in purely photon-counting mode relative to an ideal purely photon-counting detector. The remaining two factors are related to spectral information. The DQE of an ideal purely photon-counting detector is 0.99 for water detection and 0.65 for iodine detection, reflecting the fact that there is more benefit to energy discrimination for iodine imaging compared to density imaging. Finally, the energy weighting factor $f_{ew}$ describes how much of the available energy information the studied detector is able to recover if it is allowed to weight the different energy bins optimally. For low-to-medium-frequency tasks, the energy information does not give much extra benefit for the water detection task (0.4-5 $\%$), but makes an important contribution for the iodine detection task (12-23 $\%$) where the signal is concentrated at low energies. For high-frequency tasks, energy weighting allows the detector to recover some of the information lost due to the $\mathrm{MTF}^2$ factor, which can be seen from the large $f_{ew}$ near the zero-crossing of the MTF. This can be understood by noting that the detector aperture size, and therefore the point where the MTF crosses zero, is different for different energy bins. By changing the weighting of the energy bins as a function of frequency, it is therefore possible to obtain a nonzero DQE at every frequency, causing the rise in $f_{ew}$.

The simulation of object scatter (Table \ref{table:SPR}) demonstrates the trade-off between geometric efficiency and scatter rejection, with denser grids giving lower SPR. As shown in table \ref{table:SPR} the optimal scatter DQE factor is obtained for 0.1 mm thick W lamellae with 1 mm spacing in a 1D grid, with equal performance with 2 mm spacing for 60 mm Si with W foils. Compared to this optimum, the performance of the 2D grid with 0.1 mm thick lamellae is only 2-5 $\%$ inferior. The additional scatter rejection of a 2D vs. a 1D grid is somewhat smaller than the loss in geometric detection efficiency due to the added lamellae. However, as can be seen in Table \ref{table:DQE_scatter_factor}, the 2D design gives 54-72 $\%$ lower SPR than the 1D design with the same lamella thickness and pitch. This suggests that using a 2D grid may be an option in situations where not only the DQE but also the absolute scatter intensity is important, e.g. when avoiding scatter artifacts is of high importance. Conversely, if a larger SPR is tolerable, it may be desirable to space the grid lamellae 2 mm apart in order to reduce manufacturing cost.

The 1D grid with 0.2 mm thick W lamellae and 1 mm pitch exhibits inferior performance to compared to the other 1D and 2D designs with 1 mm pitch. Increasing the lamella thickness from 0.1 mm to 0.2 mm decreases the SPR by at most one percentage point which is not enough to outweigh the penalty in geometric detection efficiency.

Table \ref{table:SPR} also shows that the 60 mm Si design with internal W foils has a substantially lower SPR for all one-dimensional external grid configurations. Since the internal foils are orthogonal to the external lamellae, they together form a 2D grid. Thus, the internal foils, originally motivated by the need to block internal scatter in the detector, have the additional benefit of rejecting scatter from the object. Consequently, the lower intrinsic geometric detection efficiency of this detector design is compensated for by the lower SPR, so that the total DQE factor from object scatter and GDE is similar for all studied designs (Table \ref{table:DQE_scatter_factor}). Correcting the DQE with this factor (Table \ref{table:DQE0_with_scatter}) therefore improves the relative performance of the Si detector with W foils compared to the intrinsic DQE in Table \ref{table:DQE0}.

Our evaluation of object scatter is based on some approximations. The DQE correction factor  $\mathrm{GDE}/(1+\mathrm{SPR})$ assumes that the primary and scattered photons have the same energy distributions, but in practice the scattered spectrum is shifted towards lower energies. This means that the actual impact of object scatter on DQE will depend on the relative importance of different energies in each particular imaging task. In particular, this effect could have an important impact on the performance for material quantification tasks which are strongly dependent on the detected spectrum. 

Comparing the zero-frequency DQE of the studied detector designs (table \ref{table:DQE0}), we see that increasing the absorption length of the Si detector from 30 to 60 mm leads to 5-20 $\%$ improvement in zero-frequency DQE. It may therefore be worthwhile to use the larger absorption length, even though cost and physical space requirements can make this challenging. Comparing the CdTe and Si designs at zero frequency with scatter and cross-talk included (Table \ref{table:DQE_scatter_factor}), we see that the performance of the CdTe system with 0.5 mm pixels is on par with or better than that of the 60 mm Si system with tungsten, with 28-41 $\%$ higher DQE for detection tasks and with 2 $\%$ lower to 11 $\%$ higher DQE for quantification tasks. The 0.225 mm pixel CdTe design also has higher detection DQE than the 60 mm Si design by 5-29 $\%$, depending on task and charge cloud model. The difference is largest for the density imaging task (water detection) where the lower detection efficiency of silicon dominates the result. For two-material decomposition, however, the 0.225 mm pixel CdTe design has 31-54 $\%$ lower DQE. The above findings reflect the higher energy resolution of the silicon system, in particular compared to the 0.225 mm system.

The advantage of CdTe compared to Si for detection tasks can potentially be diminished if detectors with lower noise are produced in the future. According to the low-noise simulation for an 1-keV lower threshold (Fig. \ref{fig:DQECurvesLowNoise}), with scatter taken into consideration, the DQE advantage over silicon for CdTe with 0.5 mm pixels is 27 $\%$ and 18 $\%$ for water and iodine respectively, and 14 $\%$ and 1 $\%$ for CdTe with 0.225 mm pixels and the large charge cloud model.

The detection and quantification performance for single- and dual-spectrum imaging with photon-counting detectors is compared in Fig. \ref{fig:DQECurvesDE_Si_andCdTe}. As shown in this figure, replacing a 120 kVp single energy acquisition with a 80/Sn140 kVp dual-spectrum acquisition can give a small dose efficiency improvement for detection of 1-10$\%$ relative to a single-energy acquisition but a large improvement of $101-244\%$ for iodine quantification in a two-material decomposition. Note that this improvement, to a large extent, is caused by the additional spectral information obtained from using two beam energies, since a system where both detectors measure the sum spectrum has an iodine quantification DQE that is only 8 or 21 $\%$ better for CdTe and 33 $\%$ worse for silicon, relative to single-energy imaging. Re-optimizing the energy bins for measuring the sum spectrum in both detectors improves the performance somewhat, by 11 $\%$ and 5 $\%$ for the 0.5 mm and 0.225 mm CdTe systems and by 28 $\%$ for Si, but this is far from closing the gap to the achievable performance with dual-spectrum imaging. This shows that measuring the transmitted spectrum under two different illumination conditions helps compensate for the imperfect spectral response of both CdTe and Si. Our findings are consistent with published experimental results from single- and dual-energy acquisitions with a CdTe-based photon-counting detector.\cite{tao_dual_source}

For iodine detection on the other hand, the relative dose efficiency is lower for the summed 80/Sn140 kVp spectrum than for the 120 kVp spectrum, and the additional dual-spectrum information is just barely able to compensate for this drawback. Also note that even an ideal detector performs worse for dual-spectrum imaging than for single-energy imaging for the iodine detection task, as the dose efficiency relative to an ideal single-energy system is 0.89. This can be compared to the water detection and iodine quantification tasks, for which the relative dose efficiency is 1.09 and 1.2, respectively. In contrast to the water detection and iodine quantification tasks, the iodine detection task depends on having a large amount of data at low energies, and this would require more output to be allocated to the low-kVp x-ray tube. Optimizing the x-ray output allocation between the two x-ray tubes when used with photon-counting detectors is out of the scope of this work but would be an interesting future research topic. Finally, the effect of object scatter, which is not included here, could be studied in a more comprehensive study taking into account the fact that object scatter depends on the kVp. The impact of cross-scatter, if present, should also be considered.


Importantly, the present study does not include the effect of pileup, i.e. the presented comparison applies to imaging in the low-count-rate limit. However, a design optimization of a photon-counting detector system cannot be made on the basis of the low-count-rate limit alone, since there is a trade-off between charge sharing, which is most severe for small pixel sizes, and pileup, which is most severe for large pixel sizes. Even though our results show that the performance of the 0.225 mm pixel CdTe detector is inferior to the 0.5 mm pixel CdTe detector in the low-count-rate limit, the former is much less susceptible to pileup. A fully exhaustive detector comparison should therefore be carried out for a range of photon fluence rates. Nonetheless, the low-count-rate limit is very important since the noise in a CT image is dominated by the noisiest projection lines, i.e. the projection lines with lowest x-ray fluence rate. The low-count-rate DQE presented here is therefore important, in particular in the central regions of large patients. The method of comparison used in this work could also be extended in the future with models of how the registered spectrum and noise correlations are affected by pileup, in order to fully model the rate dependent energy response.

When comparing the silicon and CdTe detectors, it is important to remember that silicon has the advantage of being easily segmentable along the depth direction; designs with 9 or 16 depth segments have been presented\cite{persson_energy_resolved_ct_imaging,liu_count_rate_performance} Even though the CdTe and Si detectors with 0.5 mm pixels receive the same number of counts per time in one pixel area, a depth segmented silicon detector could therefore be less sensitive to pileup than the CdTe detector. Since the present study does not include pileup, we have assumed that the silicon detector is not depth segmented. Introducing depth segmentation would lead to increased charge sharing, although less so than in a two-dimensional detector since the depth segments are typically longer than the transverse pixel size. A future extension of the present framework to include pileup could be used to investigate the impact of depth segmentation and how the number of depth segments should be chosen to give a favorable trade-off between charge sharing and count rate. Another future topic of investigation could be the impact of anti-coincidence logic in CdTe and Si detectors to mitigate the detrimental effects of charge sharing.

A subtle source of error can affect the results at high spatial frequencies. In our simulation, we sampled the detector point-spread function with three subsamples per pixel. Although the effect of the sub-beam size on the MTF is corrected by Fourier division, the discrete sampling also means that the simulated MTF is an aliased version of the true MTF, with the replicas spaced by three times the sample frequency. This causes an error in the estimated MTF that becomes important for frequencies near and above the sampling frequency. Based on the magnitude of this phenomenon for an idealized pixel with no cross-talk, we expect that this could overestimate the MTF by about 4\% at the Nyquist frequency and about 25\% near the sampling frequency, where the DQE is near 0, making this difference hard to discern visually. If one wants to study the exact behavior of the DQE and its constituent factors at or above the sampling frequency, it may therefore be necessary to perform a more extensive simulation with more sub-beams.

Finally, the simulations presented here focused on the performance limitations of the detector itself and did not take the effect of the x-ray focal spot into account. In practice, photon-counting systems with small pixels are likely to be resolution limited by the focal spot size, so that the resolution difference between 0.5 and 0.225 mm pixel size is smaller in practice than suggested by Fig. \ref{fig:DQECurvesCdTe}. Extending the simulation procedure used here in order to investigate the system performance taking both focal spot and detector properties into account is a topic for future research.

\section{Conclusion}
In this work, we demonstrated how a linear-systems framework can be used to make quantitative comparisons of different photon-counting CT detector designs, taking spatioenergetic detector imperfections into account. We also demonstrated realistic models for charge sharing in CdTe and Si, fitted to previously published data. Furthermore, we demonstrated a way of breaking down the DQE into a product of factors which can be useful for understanding the impact of different physical factors on detector performance.

Our results show that the CdTe detectors, in particular the design with 0.5 mm pixels, outperforms the studied silicon designs for detection of water and iodine, whereas the 60 mm silicon system outperforms the 0.225 mm CdTe system for two-material decomposition and the 0.5 mm CdTe system for three-material decomposition. Together with the fact that silicon detectors are easily depth segmentable, potentially able to reject object scatter with internal blockers and the material is readily available at a reasonable cost, this suggests that Si should be considered for photon-counting CT. We also demonstrated that dual 80/Sn140 kVp combined with photon counting can give a large improvement in dose efficiency over single-energy for photon-counting imaging with either CdTe or Si. Further investigations will be necessary to investigate the effect of focal spot blur and pileup and the possibility of countering pileup by depth segmenting the Si detector.
\label{sec:conclusion}

\subsection*{Disclosures}
Mats Persson is presently a visiting researcher with General Electric Company, with funding from the EU Research Executive Agency. Mats Persson is stockholder of and consultant for Prismatic Sensors AB. Norbert J. Pelc is consultant for Prismatic Sensors AB.

\acknowledgments 
Our laboratory receives research funding from GE Healthcare and NIH (U01 EB 017140). We thank Cheng Xu and Xuejin Liu for providing the measured spectra used to fit the charge sharing models, and Fredrik Gr\"{o}nberg and Yifan Zheng for helpful discussions. Some of the computing for this project was performed on the Sherlock cluster at Stanford. We would like to thank Stanford University and the Stanford Research Computing Center for providing computational resources and support that contributed to these research results. Computations were also performed on resources provided by the Swedish National Infrastructure for Computing (SNIC) at PDC.


\vspace{2ex}\noindent\textbf{Mats Persson} received his MSc degree in Engineering Physics in 2011 and his PhD degree in Physics in 2016, both from KTH Royal Institute of Technology, Stockholm, Sweden. After receiving his PhD he has worked as a postdoctoral researcher at Stanford University and is presently a visiting postdoctoral researcher at GE Research Center, Niskayuna, NY. His research interests include detector development, performance modeling and data processing methods for photon-counting spectral CT.

\vspace{2ex}\noindent\textbf{Adam Wang} is an Assistant Professor of Radiology at Stanford University. He completed his PhD in 2012 from Stanford and his postdoctoral fellowship in 2014 at Johns Hopkins University, followed by several years (2014-2018) at Varian Medical Systems as a senior scientist. His research interests include novel designs for x-ray and CT systems, spectral x-ray imaging, and reconstruction algorithms. 

\vspace{2ex}\noindent\textbf{Norbert J. Pelc} received the B.S. degree in applied mathematics, engineering and physics from the University of Wisconsin-Madison in 1974, and the M.S. and D.Sc. degrees in medical radiological physics from Harvard University in 1976 and 1979. He is currently Boston Scientific Applied Biomedical Engineering Professor and Professor of Radiology, Emeritus at Stanford University. He has authored over 200 peer-reviewed papers and over 340 conference presentations and is the inventor of 95 issued U.S. patents.
\vspace{1ex}

{\listoffigures}
{\listoftables}

\end{spacing}
\end{document}